\documentclass[aps,pra,reprint,superscriptaddress]{revtex4-2}

\usepackage{graphicx}
\usepackage[normalem]{ulem}
\usepackage{amsmath}
\usepackage{bbold}
\usepackage{color}
\usepackage[linktocpage=true]{hyperref}
\hypersetup{
    colorlinks,
    citecolor=blue,
    filecolor=blue,
    linkcolor=blue,
    urlcolor=blue
}

\begin{document}

\title{Single-shot determination of quantum phases via continuous measurements}

\author{Aniket Patra}
\affiliation{Max-Planck-Institut f\"{u}r Physik komplexer Systeme, D-01187 Dresden, Germany}
\affiliation{Department of Physics and Astronomy, Aarhus University, DK-8000 Aarhus C, Denmark}

\author{Lukas F. Buchmann}
\affiliation{Department of Physics and Astronomy, Aarhus University, DK-8000 Aarhus C, Denmark}

\author{Felix Motzoi}
\affiliation{Department of Physics and Astronomy, Aarhus University, DK-8000 Aarhus C, Denmark}
\affiliation{Forschungszentrum J\"{u}lich, Institute of Quantum Control (PGI-8), D-52425 J\"{u}lich, Germany}

\author{Klaus M{\o}lmer}
\affiliation{Department of Physics and Astronomy, Aarhus University, DK-8000 Aarhus C, Denmark}

\author{Jacob Sherson}
\affiliation{Department of Physics and Astronomy, Aarhus University, DK-8000 Aarhus C, Denmark}

\author{Anne E. B. Nielsen}
\affiliation{Max-Planck-Institut f\"{u}r Physik komplexer Systeme, D-01187 Dresden, Germany}
\affiliation{Department of Physics and Astronomy, Aarhus University, DK-8000 Aarhus C, Denmark}

\begin{abstract}
We propose that weak continuous probing may be exploited to determine and define quantum phases of complex many-body systems based on the measurement record alone. We test the resulting phase criterion in numerical simulations of measurements on the Bose-Hubbard model and the quantum Ising chain. This yields a phase transition point in reasonable agreement with the quantum phase transition in the ground state of the closed system in the thermodynamic limit, despite the system being highly excited through the measurement dynamics. At high measurement strengths, the system's response enters a Zeno regime suppressing transitions between eigenstates of the measurement operator.
\end{abstract}

\maketitle

\section{Introduction}

Quantum phases allow descriptions of complex systems in simpler terms than a microscopic description \cite{sachdev}. Distinct phases span wide areas in parameter space characterized by the fundamental excitations, which govern the system's equilibrium properties and response to perturbations.  They are used to characterize a wide range of physical phenomena, including electronic, magnetic and optical properties of solid state systems \cite{review0, elsystems}, nuclear physics \cite{nuclear} and cosmological topological defects \cite{kibble,zurek}.

\begin{figure}
\includegraphics[trim={5cm 7cm 3cm 0cm},clip, scale=0.33]{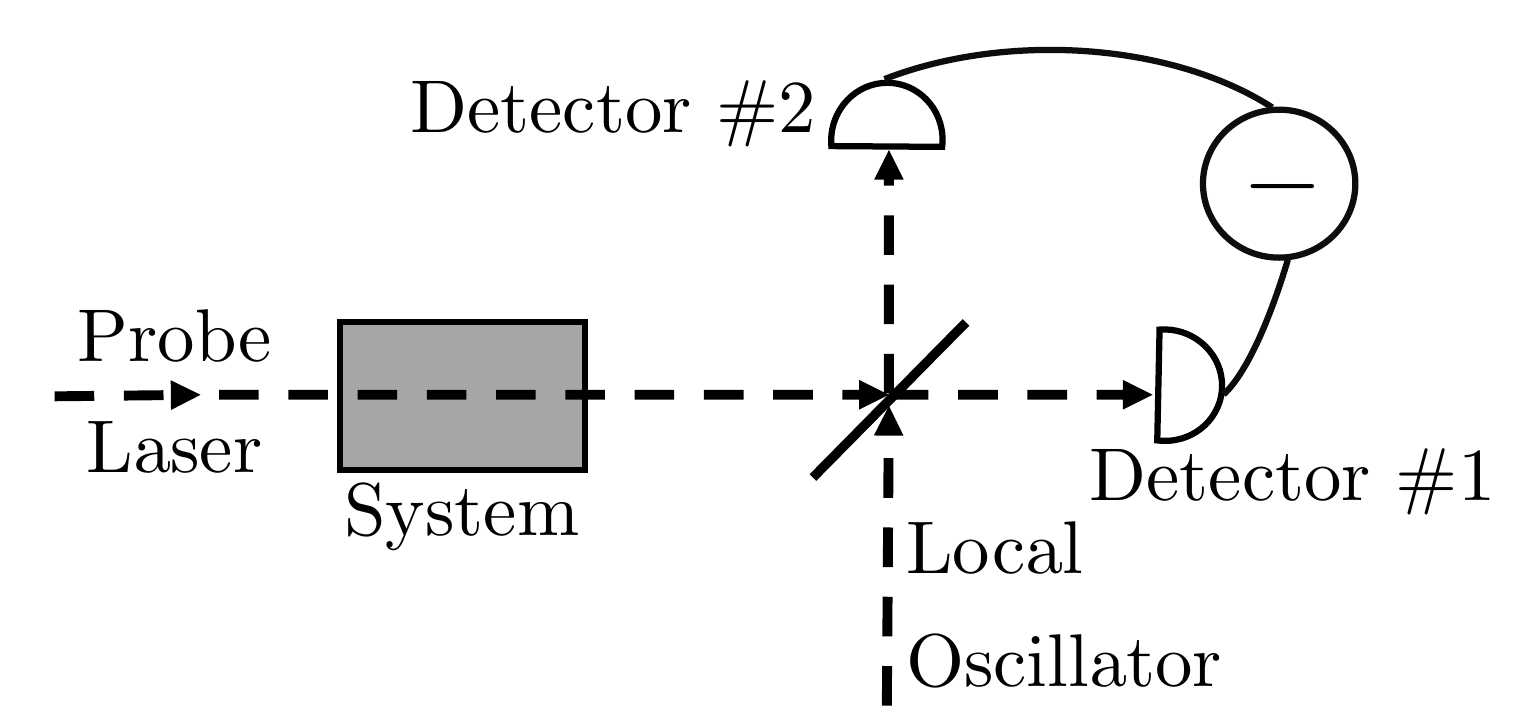}
\caption{The experimental setup shows a balanced homodyne measurement setup, realizing the signal and dynamics represented by Eqs.\ \eqref{eq:homomeasure} and (\ref{eq:homodyne}).  The cavity (the gray shaded box) contains strongly interacting Hamiltonians engineered using cold atoms in optical lattices.  We depict the 50:50 beam splitter as the slanted straight line. }
\label{HomodynePic}
\end{figure}

We consider a closed system described by a Hamiltonian
\begin{equation}
\hat{\mathcal{H}} = \hat{\mathcal{H}}_0 + \alpha \hat{\mathcal{H}}_1 \label{eq:HamiltonAlpha}
\end{equation}
that exhibits a single quantum phase transition with an abrupt change in the order parameter at the critical value $\alpha = \alpha_c$. The change is often due to an avoided crossing in the energy spectrum, and the location is uncovered by a change of the ground state expectation value of an appropriate order parameter that becomes infinitely sharp in the thermodynamic limit.

In dynamical phase transitions \cite{diehl2010dynamical, heyl2013dynamical, weimer2015dynamical, Patra1, Patra2, Patra3}, the situation is considerably richer than the exploration of ground state properties since the entire spectrum contributes to the dynamics. This leads to intricate questions about excited state phase transitions \cite{perez2011quantum,stransky2016classification} and accidental dynamical phase transitions \cite{vajna2015topological}. Rather than suddenly changing the Hamiltonian, another way to quench a quantum system is to perform a measurement \cite{quench,klaus1,jjZeno,fumes}.

When studying phase transitions experimentally, a measurement typically destroys the system. One performs the experiment several times to acquire signal or statistics. Such averaging introduces variances into otherwise well-defined parameters, e.g., particle number \cite{particlenumbers}.  Although modifications of complex systems by measurement have been studied \cite{Minganti,ueda, mekhov, moreno}, the fundamental question -- if the quantum phase of a complex system can be determined from the measurement record alone -- remains, to the best of our knowledge, unanswered.

In this article,  we propose a criterion for the determination of quantum phases based solely on the measurement record of a single experimental run. This proposal relies on continuous dispersive measurements.  It is well known that even weak and continuous measurements induce a back action through the noisy measurement record that builds up to a substantial perturbation of the system \cite{klaus1}.  Here, we exploit this to disturb the system and simultaneously record its response. Similar to dynamical phase transitions, the entire spectrum contributes to the response. The measurement strength sets the magnitude of the disturbance. At low measurement strength, we numerically demonstrate that we can extract information about the system's phase transition.

After introducing our criterion, we apply it to the Bose-Hubbard model and the quantum Ising chain.  We show that our criterion agrees reasonably with the known phase transition in the thermodynamic limit, despite the system is not in the ground state. We also demonstrate how the measurement strength itself becomes a parameter in the open system's phase diagram revealing (potentially controllable) properties of strongly probed systems.

\section{Phase Determination}

Let $\hat{\mathcal{M}}_0$ be a Hermitian operator satisfying $[\hat{\mathcal{H}_0},\hat{\mathcal{M}}_0]=0$ and [$\hat{\mathcal{H}_1},\hat{\mathcal{M}}_0]\neq0,$ where $\hat{\mathcal{H}}_0$ and $\hat{\mathcal{H}}_1$ are the non-commuting Hamiltonians in Eq.~\eqref{eq:HamiltonAlpha}. Consider a probe that dispersively measures $\hat{\mathcal{M}}_0$ with strength $\gamma$. The probe yields a measurement record $I(t)$ and disturbs the system through the measurement back-action. The experimental setup is shown in Fig.\ \ref{HomodynePic}.

For concreteness, we assume a homodyne measurement signal given by
\begin{equation}
I(t) = 2\gamma \langle \hat{\mathcal{M}_0} \rangle + \sqrt{\gamma} \: \text{d}W/\text{d}t
\label{eq:homomeasure}
\end{equation}
 where $\langle \cdot \rangle$ denotes the expectation value and $\text{d}W$ is a Wiener increment. The state of the system conditioned on the measurement outcome evolves according to the It\^{o} stochastic Schr\"{o}dinger equation (SSE)
\begin{equation}
\text{d}|\bar{\psi}(t)\rangle =\Bigl[-i\hat{\mathcal{H}} - \frac{\gamma}{2} \hat{\mathcal{M}}_0^2 + I(t) \hat{\mathcal{M}}_0 \Bigr]\text{d}t|\bar{\psi}(t)\rangle, \label{eq:homodyne}
\end{equation}
where $\hbar = 1$ and $|\bar{\psi}\rangle$ denotes a non-normalized state \cite{jjZeno,wiseman2009quantum,jacobsBook,jacobsIntro}. The first term in Eq.\ \eqref{eq:homodyne} describes the unitary evolution. The second and third terms include the dissipation associated with the measurement.  This approach is experimentally appealing, since it allows extracting phase information from a single continuous measurement.

Here we do a simulation of such an experiment.  We calculate the power spectral density (PSD)
\begin{equation}
S(\omega)=(2\pi T)^{-1}\mathbb{E}\bigl[|\int_0^T e^{-i \omega t} I(t)\text{d}t|^2\bigr]
\label{eq:SexpressionSSE}
\end{equation}
by numerically integrating the SSE, see Appendices \ref{SecSpecNum} and \ref{SecNumIntSSE}.  In order to sample the typical behavior away from the initial state, we discard the initial part of the quantum trajectories. We divide the considered quantum trajectory into several parts and calculate the average PSD to obtain the noise average $\mathbb{E}$.

The average dynamics, on the other hand, over different Wiener increments with $\text{d}W^2=\text{d}t$ is given by the Gorini-Kossakowski-Lindblad-Sudarshan (GKLS) master equation $\dot{\rho}= \mathcal{L}[\rho] = - i [\hat{\mathcal{H}},\rho]+\gamma\mathcal{D}[\hat{\mathcal{M}}_0]\rho$ with $\mathcal{D}[\hat{O}]\rho = \hat{O}\rho \hat{O}^{\dagger} -\frac{1}{2}\left\lbrace \hat{O}^{\dagger}\hat{O}, \rho\right\rbrace$ \cite{suppGKS, suppLindblad, suppBreuer, wiseman2009quantum, jacobsBook, jacobsIntro}. Our goal is to relate the phase properties of the system to the measurement signal's autocorrelation function $F^{(1)}_\textrm{hom}(t,t+\tau)=\mathbb{E}[I(t)I(t+\tau)]$ where $\mathbb{E}$ denotes a classical expectation value over the noise realizations. This correlation is given by the quantum regression theorem as $F^{(1)}_\textrm{hom}(t,t+\tau) = 2\gamma^2\text{ Tr}\left[\hat{\mathcal{M}}_0 e^{\mathcal{L}\tau}\left\lbrace \hat{\mathcal{M}}_0, \rho^{\text{st}}\right\rbrace\right]$, where $\rho^{\text{st}}$ is the stationary state, such that $\mathcal{L}[\rho^{\text{st}}]=0$ \cite{gardiner2004quantum}, see also Appendix \ref{SecF1} for details. The identity is always a stationary state since $\hat{\mathcal{M}}_0$ is Hermitian.  To verify this, one replaces $\rho^{\text{st}} = \mathbb{1}/N$ in the GKLS master equation and uses $\mathcal{D}[\hat{\mathcal{M}}_0]\mathbb{1} = 0$. Here $N$ is the dimension of the Hilbert space.  This can be understood as the measurement back-action acting as an infinite temperature heat-bath in the long-time limit \cite{optomechanics, jaksch}.

Considering $\rho^{\text{st}} = \mathbb{1}/N$,  the stationarity of the noise process, and making use of the quantum regression theorem, we obtain
\begin{multline}
S(\omega) = \frac{4\gamma^2}{N} \int_{-\infty}^\infty \text{Tr} [\hat{\mathcal{M}}_0 e^{\mathcal{L}\tau} \hat{\mathcal{M}}_0] e^{-i \omega \tau} \text{d}\tau \\
= \frac{8\gamma^2}{N} \mathrm{Re} \left[\int_0^\infty \text{Tr} [\hat{\mathcal{M}}_0 e^{\mathcal{L}\tau} \hat{\mathcal{M}}_0] e^{-i \omega \tau} \text{d}\tau\right], \label{eq:Sexpression}
\end{multline}
where $\mathrm{Re}$ is the real part. The front factor is particular to homodyne measurements \cite{wiseman2009quantum}, see also Appendices \ref{SecF1} and \ref{SecSpecAnalytical} for details.  Since $\mathcal{D}[\hat{\mathcal{M}}_0]\hat{\mathcal{M}}_0 = 0$, we have
\begin{equation}
\mathcal{L}[\hat{\mathcal{M}}_0] = - i [\hat{\mathcal{H}},\hat{\mathcal{M}}_0].
\label{CommCriterion}
\end{equation}
After expanding $e^{\mathcal{L}\tau}$ in Eq.\ \eqref{eq:Sexpression} and utilizing Eq.\ \eqref{CommCriterion}, we conclude that the PSD is determined by the commutation relations of $\hat{\mathcal{H}}$ and $\hat{\mathcal{M}}_0$.

Assuming $\mathcal{L}$ is diagonalizable with eigenvalues $\lambda_m$ and right (left) eigenmatrices $r_m \;(l_m)$, Eq.\ \eqref{eq:Sexpression} is decomposed as $S(\omega) = S_d(\omega)+S_0(\omega)$ with
\begin{subequations}
\begin{equation}
S_d(\omega) = \frac{8\gamma^2}{N}\mkern-12mu \sum_{\mathrm{Re} (\lambda_m)<0}\mkern-24mu \frac{-\mathrm{Re} (\lambda_m) \mathrm{Re} (t_m) + \left[\omega - \mathrm{Im} (\lambda_m)\right]\mathrm{Im} (t_m)}{\left[\omega - \mathrm{Im} (\lambda_m)\right]^2 + \left[\mathrm{Re} (\lambda_m)\right]^2} \label{eq:eigenPSD},
\end{equation}
\begin{multline}
S_0(\omega) = \frac{8\gamma^2}{N} \sum_{\mathrm{Re} (\lambda_m) = 0} \bigg[\pi \mathrm{Re} (t_m)  \delta\left(\omega - \mathrm{Im} (\lambda_m)\right) \\
+ \mathcal{P} \left(\frac{\mathrm{Im} (t_m)}{\omega - \mathrm{Im} (\lambda_m)}\right)\bigg], \label{eq:eigenPSDzero}
\end{multline}
\end{subequations}
where $t_m = \text{Tr}[\hat{\mathcal{M}}_0 r_m]\text{Tr}[l_m^\dagger \hat{\mathcal{M}}_0]$, $\mathrm{Im}$ the imaginary part, $\mathcal{P}$ the Cauchy principal value, and $\delta$ the Dirac-delta function, see Appendix \ref{SecSpecAnalytical}.  Here $S_d(\omega)$ ($S_0(\omega)$) is the part from all of the decaying (decay-free) eigenvalues of $\mathcal{L}$. We observe that the peaks in the spectra will be located at $\mathrm{Im} (\lambda_m)$. If $\mathrm{Im} (t_m)=0$, the eigenvalue $\lambda_m$ will contribute with a Lorentzian to the spectrum.  The eigenmatrices with non-vanishing $\mathrm{Im} (t_m)$ give rise to non-Lorentzian contributions in the spectrum.

\begin{figure*}
\includegraphics[trim={0.25cm 3cm 0.25cm 9cm},clip, width=0.5\textwidth]{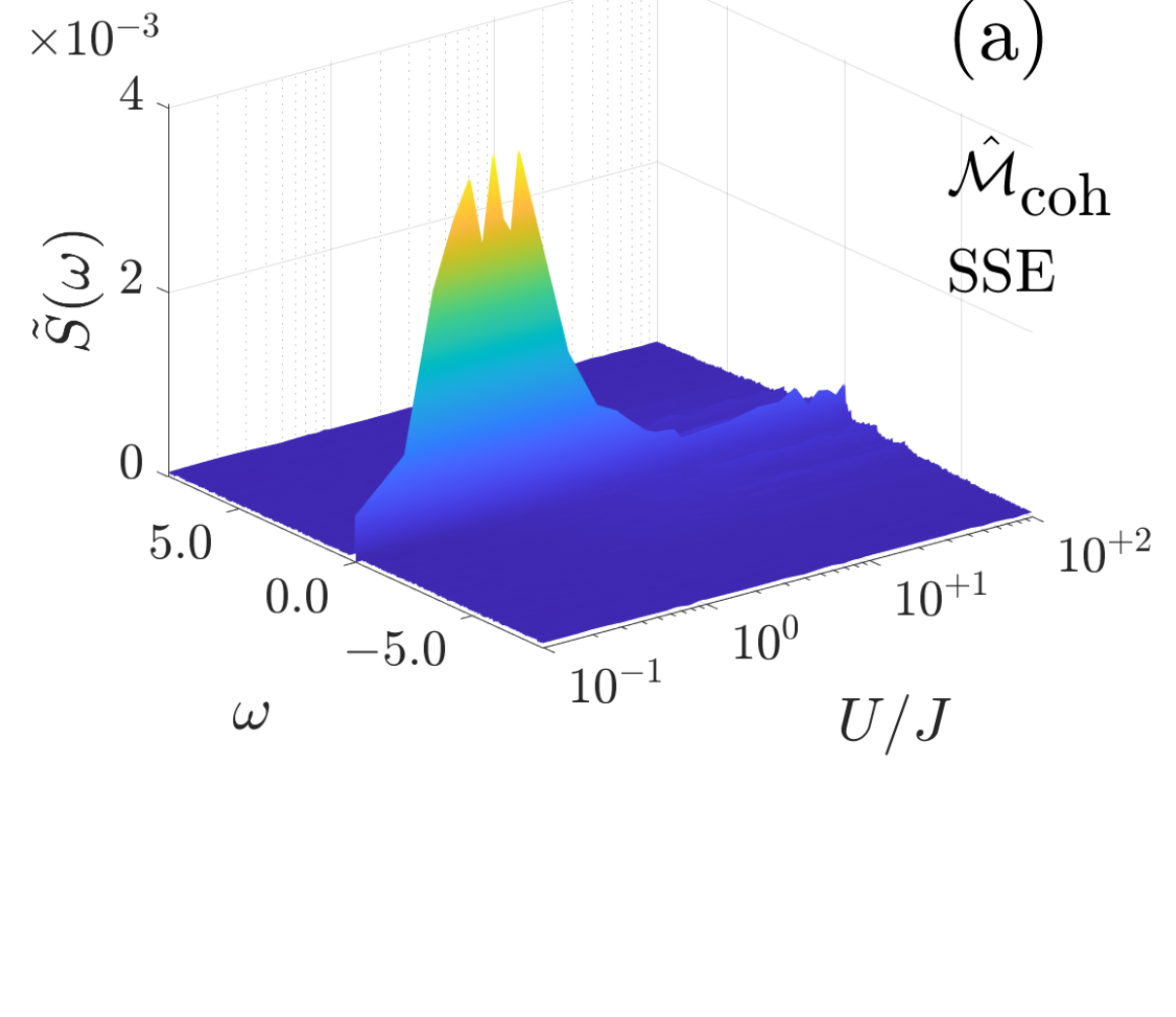}\hfill
\includegraphics[trim={0.25cm 3cm 0.25cm 9cm},clip, width=0.5\textwidth]{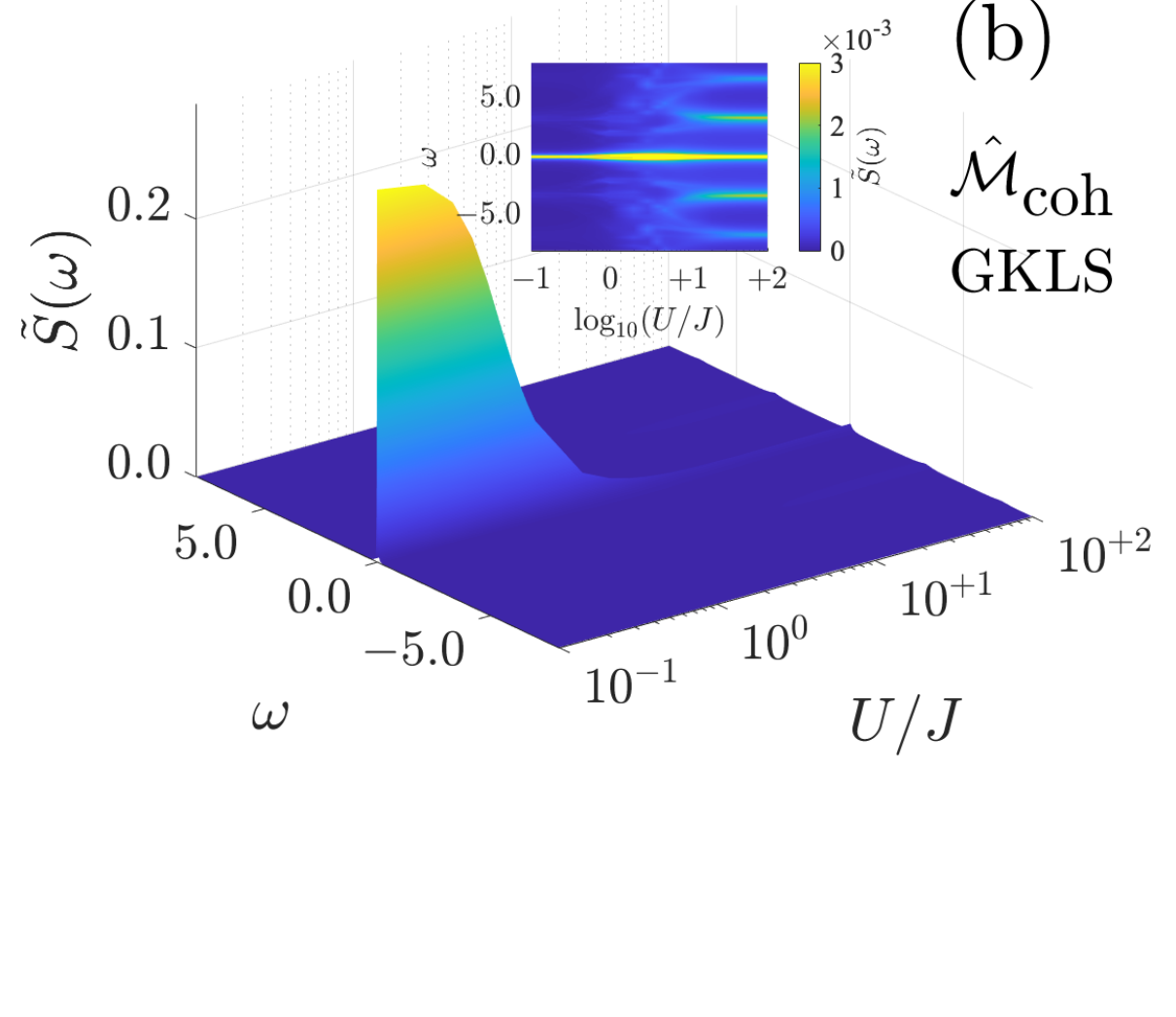}\\
\includegraphics[trim={0.25cm 3cm 0.25cm 9cm},clip, width=0.5\textwidth]{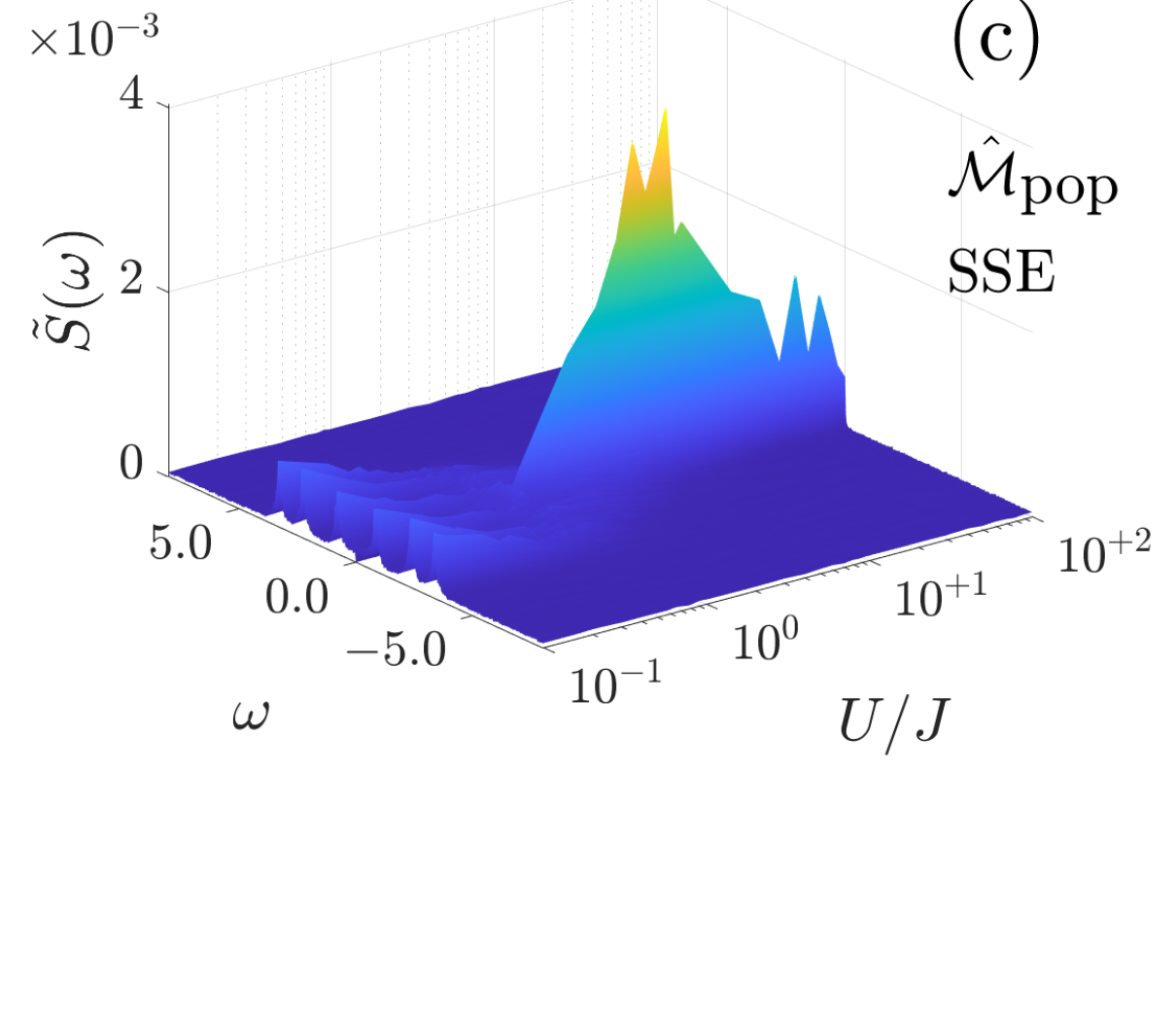}\hfill
\includegraphics[trim={0.25cm 3cm 0.25cm 9cm},clip, width=0.5\textwidth]{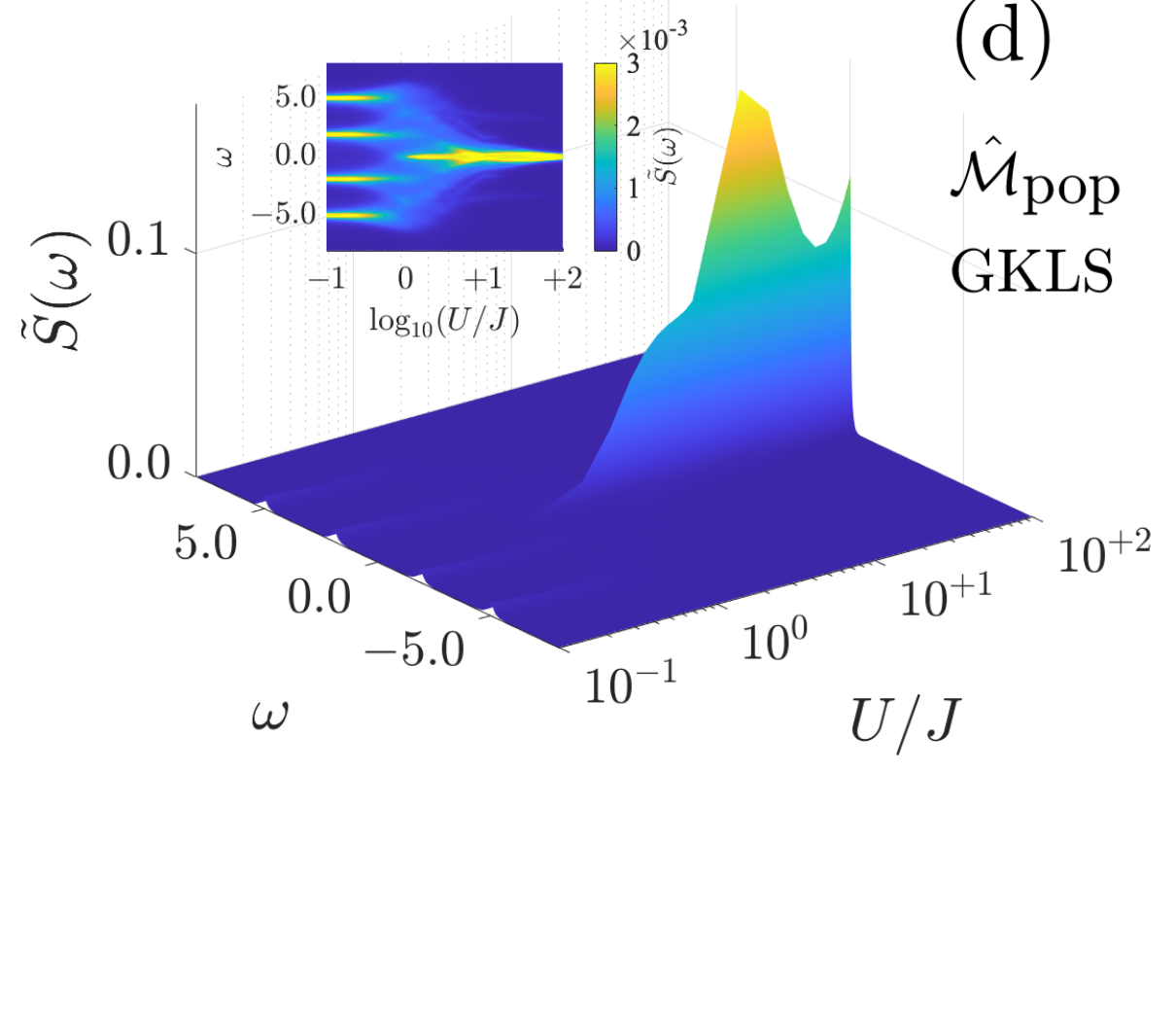}
\caption{We plot normalized PSDs $\tilde{S}(\omega)$ for the Bose-Hubbard model, where we measure two observables: $\hat{\mathcal{M}}_\mathrm{pop}=m_\mathrm{pop}\sum_{j}\hat{b}_{2j}^\dagger \hat{b}_{2j}$ and $\hat{\mathcal{M}}_\mathrm{coh}=m_\mathrm{coh}\sum_j\hat{b}_j^\dagger \hat{b}_{j+1}+h.c.$ The spectra in (\textbf{a}) and (\textbf{c}) are obtained by simulating the SSE (indicated as `SSE' in the figures), whereas (\textbf{b}) and (\textbf{d}) (indicated as `GKLS') are obtained with Eq.\ \eqref{eq:Sexpression}, where $\mathcal{L}$ is the Liouvillian appearing in the GKLS master equation -- see also Appendices \ref{SecF1} and \ref{SecSpecAnalytical} for details.  For the SSE calculations, we considered $6$ sites with $6$ particles and measurement strength $\gamma = 0.01$ for the measurement of $\hat{\mathcal{M}}_{\text{coh}}$ and $\gamma = 0.1$ for the measurement of $\hat{\mathcal{M}}_{\text{pop}}$.  The computations based on the master equation are harder to do and hence we have used $4$ sites and $4$ particles and kept the same $\gamma$ values in (\textbf{b}) and (\textbf{d}). The figures shown as insets into (\textbf{b}) and (\textbf{d}) are rotated versions of the main plots and show the abrupt change near $1 < U/J < 10$. Compare this with the order parameter vs $U/J$ plot in Fig.\ \ref{FigBHIsingOP}(a). }\label{BHMeasureFig}
\end{figure*}

As we change the parameter $\alpha$, the system (\ref{eq:HamiltonAlpha}) undergoes a phase transition at $\alpha = \alpha_c$. For second order quantum phase transitions, this is attributed to the level crossings at $\alpha = \alpha_{c}$ \cite{sachdev}. If $\gamma=0$, the unperturbed $\mathcal{L}$ has eigenvalues and vectors $\lambda_{ij}=-i(E_i-E_j)=-i\omega_{ij}$ and $r_m=l_m = |\psi_i\rangle\langle \psi_j|,$ where $\hat{\mathcal{H}}|\psi_i \rangle = E_i |\psi_i\rangle$.  For a weakly probed system, the eigenvalues and eigenvectors of $\mathcal{L}$ are obtained perturbatively.  Therefore, the PSDs -- which are related to the level statistics via $\lambda_{ij}, r_{m},$ and $l_{m}$, cf. Eqs.\ \eqref{eq:eigenPSD} and (\ref{eq:eigenPSDzero}) -- for the two different phases are also qualitatively different.  Using this, one can identify the two distinct phases in Figs.\ \ref{BHMeasureFig} and \ref{IsingMeasureFig}. Note, we have plotted the normalized PSD  $\tilde{S}(\omega) = S(\omega)/\int_{-\infty}^\infty S(\omega)\text{d}\omega$ in the aforementioned panels for numerical convenience.

\begin{figure*}
\includegraphics[trim={0.25cm 7cm 0.25cm 7cm},clip, width=0.5\textwidth]{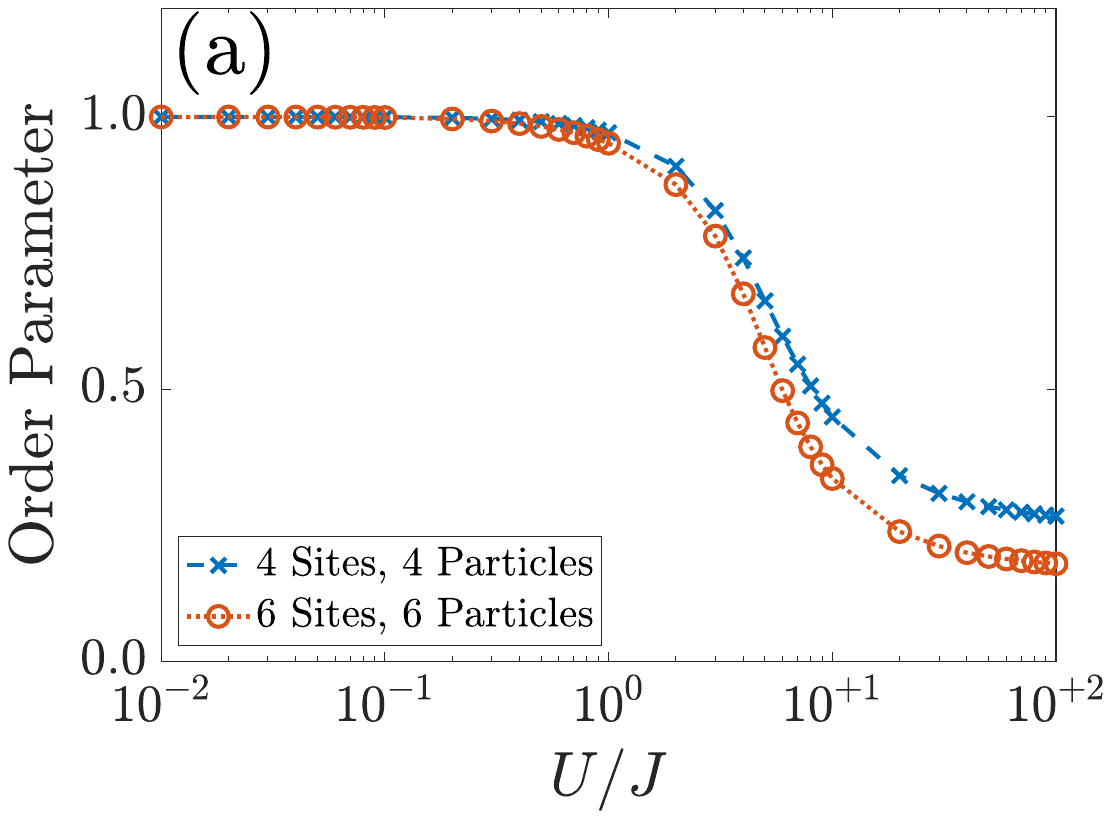}\hfill
\includegraphics[trim={0.25cm 7cm 0.25cm 7cm},clip, width=0.5\textwidth]{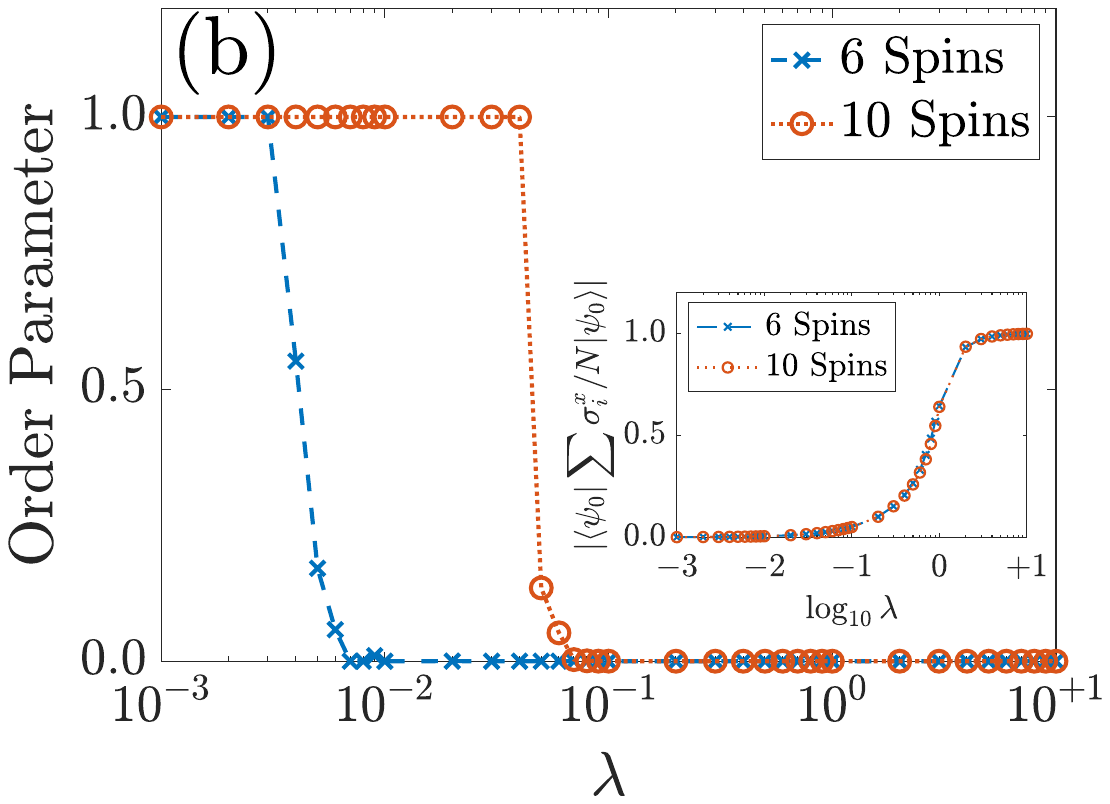}
\caption{We show the order parameters vs the parameter $\alpha$ in the Hamiltonian \eqref{eq:HamiltonAlpha} for the Bose-Hubbard and the transverse-field Ising model in Figs.\ (\textbf{a}) and (\textbf{b}), respectively.  The order parameter for the transverse-field Ising model is $|\langle \psi_0 | \sum_{i}\sigma_{i}^{z}/N | \psi_0 \rangle|$ with $| \psi_0 \rangle$ being the ground state, whereas for the Bose-Hubbard model we plot the condensate fraction \eqref{BHOPDef}.  For the Bose-Hubbard model in (\textbf{a}) the parameter $\alpha$ is the ratio between the interaction strength $U$ and the hopping strength $J$, whereas $\alpha$ is the transverse-field strength $\lambda$ for the transverse-field Ising model in (\textbf{b}). We have also included the plot of $|\langle \psi_0 | \sum_{i}\sigma_{i}^{x}/N | \psi_0 \rangle|$ vs $\lambda$ as an inset of (\textbf{b}). Since this is not the order parameter of the transverse-field Ising model,  it does not show any abrupt change similar to the main plot in panel (\textbf{b}).}
\label{FigBHIsingOP}
\end{figure*}

One can use two different choices of  measurements -- $\hat{\mathcal{M}}_0$ and $\hat{\mathcal{M}}_1$ -- that commute with different parts of $\hat{\mathcal{H}}$.  The PSD corresponding to one phase for the $\hat{\mathcal{M}}_0$ measurement is qualitatively similar to the PSD corresponding to the other phase for the $\hat{\mathcal{M}}_1$ measurement.  In particular, we obtain from Eq.\ \eqref{eq:Sexpression} that $S(\omega) \propto \delta(\omega)$ when the measured operator commutes with the Hamiltonian.

The criterion for determining a phase transition, therefore, is detecting changes in the PSD corresponding to a particular measurement $\hat{\mathcal{M}}_i$. Using different $\hat{\mathcal{M}}_i$, one can also detect multiple phase transitions.  Assuming $n_{\alpha}$ phase transitions in a Hamiltonian $\hat{\mathcal{H}}(\alpha)$, one obtains representative Hamiltonians $\hat{\mathcal{H}}(\alpha_i)$ with $i = 1,2, \ldots, n_{\alpha}+1$, where $\alpha_i$ is a parameter value corresponding to a particular phase. We consider $n_{\alpha}+1$ distinct measuring operators satisfying $[\hat{\mathcal{H}}(\alpha_i),\hat{\mathcal{M}}_i] = 0$. Note that since $[\hat{\mathcal{M}}_i,\hat{\mathcal{H}}(\alpha)] \neq 0$, our continuous measurement scheme is \textit{not} a quantum non-demolition measurement \cite{QND1, QND2}.

In the following, we implement this scheme to study the phase transitions in an ergodic (Bose-Hubbard) system and in an integrable (transverse-field Ising chain) Hamiltonian.

\section{Probed Bose-Hubbard Model}
\label{SecBHMainText}

The 1D Bose-Hubbard model provided the first demonstration of a quantum phase transition in ultracold atoms \cite{Greiner2002}, and it is a powerful tool for the experimental study of quantum phases \cite{tonks, feshbachrestuning, simon, review1, review2}, including studies of driven-dissipative quantum systems \cite{mekhov, ueda, drivendiss}. The Hamiltonian reads
\begin{align}
\hat{\mathcal{H}}=-J\sum_{\langle j,k\rangle}(\hat{b}_j^\dag\hat{b}_k+\hat{b}_k^\dag\hat{b}_j)+\frac{U}{2}\sum_j\hat{b}_j^\dag\hat{b}_j(\hat{b}_j^\dag\hat{b}_j-1), \label{eq:bhhamiltonian}
\end{align}
where the bosonic field operators are expanded in Wannier functions $\hat\Psi(x,t)=\sum_j\hat{b}_j(t)w_j(x)$, and $J$ and $U$ are the hopping and the on-site interaction, respectively. For $\alpha = U/J$ below the critical value, the system's ground state exhibits long range phase-coherence and it is a superfluid. Above that critical value, the ground
state features Fock-states on each
site and the system is in the Mott-insulator phase.

Let us now dispersively probe this system with an optical cavity field aligned with the trapping lattice. The probe light is described as $\hat{a}(t)f_a({\bf x}, \omega_\mathrm{L})e^{-i\omega_\mathrm{L}t} $ with $\omega_\mathrm{L}$ the probe frequency and $f_a({\bf x}, \omega_\mathrm{L})$ the spatial mode function. Here we treat the system in 1D.  For a Fabry-P\'erot cavity, we have $f_a(x,\omega_\mathrm{L})\propto\cos(k_\mathrm{L}x)$ with $k_\text{L}$ being the wavenumber for the probe light.

We focus on two relevant cases, namely where the probe has twice the period of the trapping potential and when the probe and the lattice have the same periodicity, but a $\pi/2$ phase shift. In the former case, this leads to a measurement operator $\hat{\mathcal{M}}_\mathrm{pop}=m_\mathrm{pop}\sum_{j}\hat{b}_{2j}^\dagger \hat{b}_{2j}$, where $m_\mathrm{pop}$ is a constant calculated from the Wannier functions, see Appendix \ref{SecImplMeasure}.  This operator commutes with the interaction term in \eqref{eq:bhhamiltonian} but not with the hopping. In the second case, we measure the sum over coherences, $\hat{\mathcal{M}}_\mathrm{coh}=m_\mathrm{coh}\sum_j\hat{b}_j^\dagger \hat{b}_{j+1}+h.c.$ (see Appendix \ref{SecImplMeasure}), which commutes with the hopping term but not the interaction.

We numerically calculate the PSDs for both $\hat{\mathcal{M}}_\mathrm{pop}$ and $\hat{\mathcal{M}}_\mathrm{coh}$. To perform the numerical integrations of the SSE (\ref{eq:homodyne}) in Fig.\ \ref{BHMeasureFig}(a,c), we considered a system with six sites and six particles. We used a system with a smaller Hilbert space -- four sites and four particles -- to obtain the PSDs using Eq.\ \eqref{eq:Sexpression} in Fig.\ \ref{BHMeasureFig}(b,d). For all the PSDs, the spectral range is rescaled to 20 in dimensionless units.  The PSDs for a particular measurement -- e.g., Fig.\ \ref{BHMeasureFig}(a,b) -- obtained from Eqs.\ \eqref{eq:homodyne} and (\ref{eq:Sexpression}) appear similar. This is because of the ergodicity of the Bose-Hubbard model.

We observe large values of the PSD at $\omega=0$ ($\omega \neq 0$) when the measurement operator and Hamiltonian is (is not) compatible with the quantum phase. The $\hat{\mathcal{M}}_\text{coh}$ PSDs in the superfluid (Mott-insulator) part is qualitatively similar to the Mott-insulator (superfluid) part of the $\hat{\mathcal{M}}_\text{pop}$ PSDs. The measurement for both operators gives the transition within the same order of magnitude, which is also in agreement with its value in the thermodynamic limit \cite{valueScalettar, valueMonien,valueKuhner}.  Additionally, we point out that the phase transition point from the PSDs are consistent with the behaviour of the order parameter (condensate fraction \cite{BHED1})
\begin{align}
f_{c} =  \lambda_1/N  \label{BHOPDef}
\end{align}
in Fig.\ \ref{FigBHIsingOP}(a), where $\lambda_1$ is the largest eigenvalue of the single-particle density matrix $\rho^{(1)}$ and $N$ is the number of particles. The matrix elements of $\rho^{(1)}$ are given by
\begin{align}
\rho^{(1)}_{ij} =  \langle \psi_0 | \hat{b}_i^\dagger \hat{b}_{j} | \psi_0 \rangle,  \label{SPDMDef}
\end{align}
where $| \psi_0 \rangle$ is the ground state of the Hamiltonian \eqref{eq:bhhamiltonian}. While comparing with the above critical value of the parameter, one needs to, however, keep in mind the significant finite size effects \cite{BHED1}, see also Fig.\ \ref{FigBHIsingOP}(a). The numerical results for the thermodynamic limit \cite{valueMonien,valueKuhner} are obtained by calculating the energy gap between the ground state and first excited state for different system sizes, and extrapolating to the infinite system.

\subsection{Strong Measurement}

\begin{figure}
\includegraphics[trim={0.25cm 3cm 0.25cm 9cm},clip, width=0.45\textwidth]{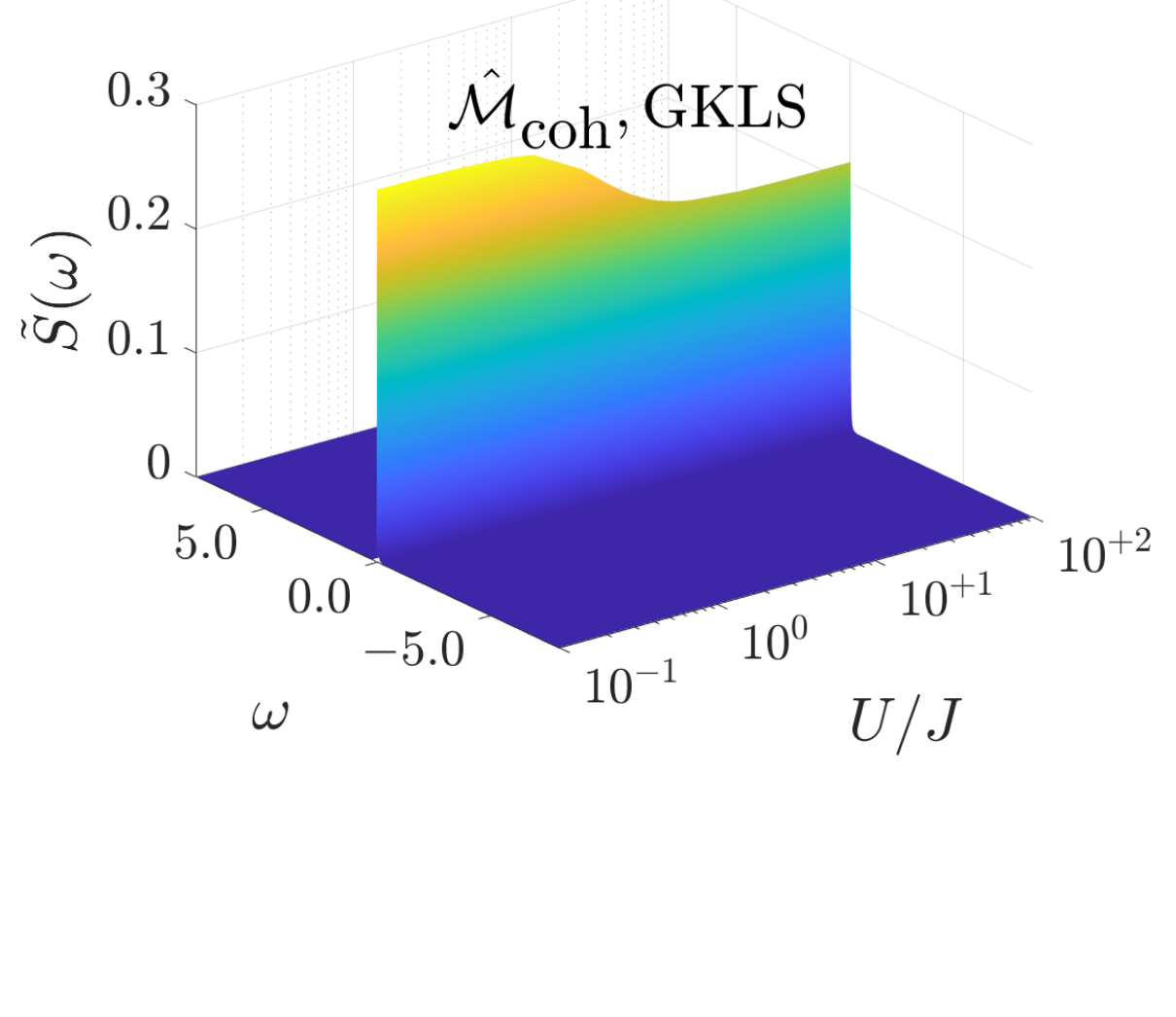}
\vspace{-2.0cm}
\caption{Quantum Zeno regime for the $\hat{\mathcal{M}}_\mathrm{coh}$ measurement in the Bose-Hubbard model.  We derive the spectrum using Eq.\ \eqref{eq:Sexpression}. When the measurement strength is very high $(\gamma = 100.0),$ the nature of the PSD does not change over a broad range of $U/J$.  This is unlike Fig.\ \ref{BHMeasureFig}(a,b). }
\label{fig:Zeno}
\end{figure}

The measured system has phase transitions defined and/or controlled by the measurement itself. That transition due to strong measurement is also witnessed by the record.  The measurement strength is considered a free parameter and an additional dimension of the phase diagram, which then depends on the operator being measured. Figure \ref{fig:Zeno} shows the PSD for measuring $\hat{\mathcal{M}}_{\text{coh}}$ with $\gamma \gg 1$ in the Bose-Hubbard model using Eq.\ \eqref{eq:Sexpression}. The figure shows that the measurement forces the system to evolve into eigenstates of the probed operator over a broad range of $U/J$.  Since $\hat{\mathcal{M}}_{\text{coh}}$ commutes with the Bose-Hubbard hopping term, the PSD implies a superfluid phase throughout. This has been identified previously as a dynamical phase transition into a Zeno regime \cite{zenoECG, zenophase, klaus2, zenoSDG}. In the given example, we demonstrate how a strong measurement of coherence turns a Mott-insulator into a superfluid. Performing strong measurements with other operators yield similar results.

\section{Probed Transverse-Field Ising Chain}
\label{SecTFIMainText}

\begin{figure*}
\includegraphics[trim={0.25cm 3cm 0.25cm 9cm},clip, width=0.5\textwidth]{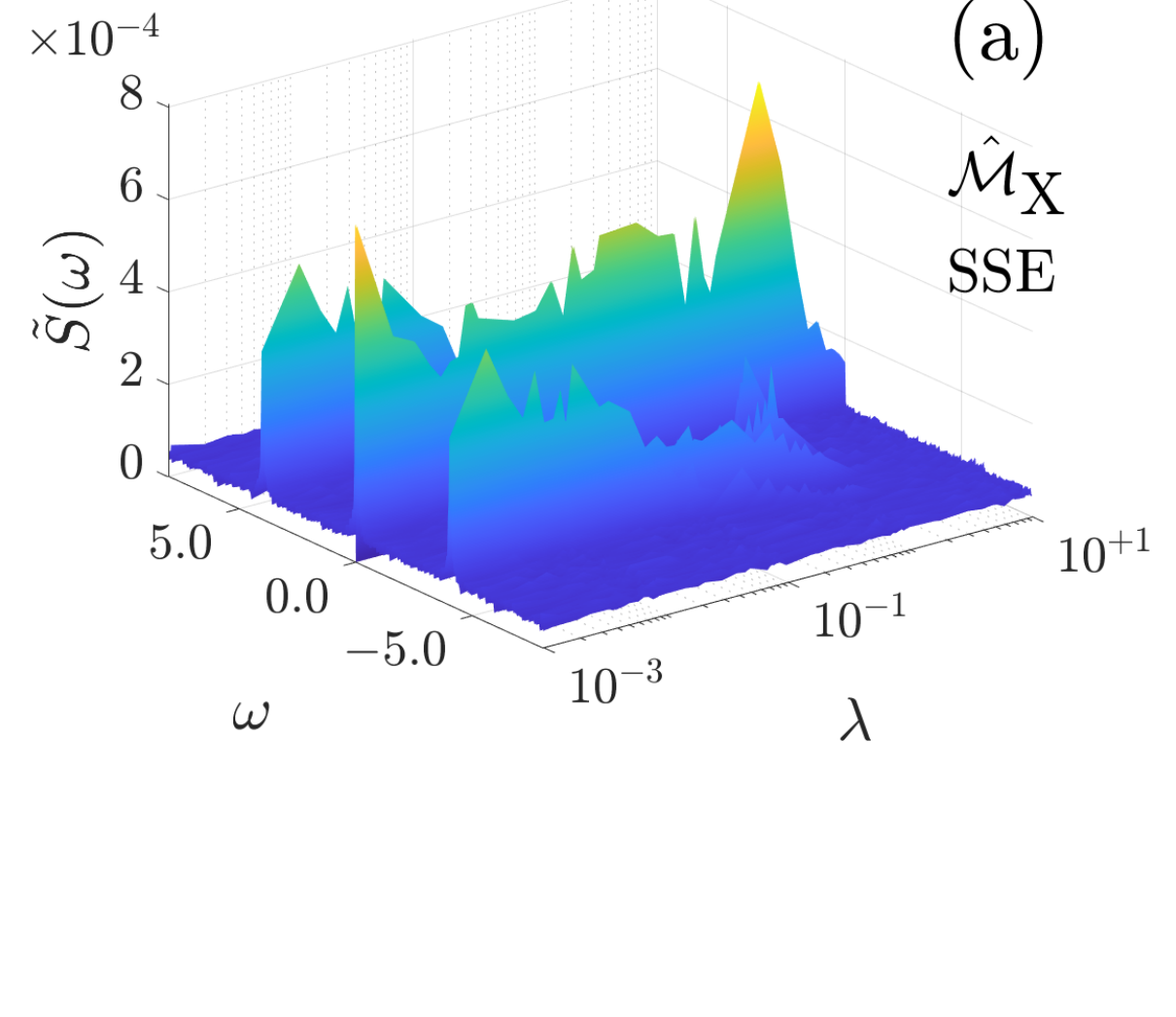}\hfill
\includegraphics[trim={0.25cm 3cm 0.25cm 9cm},clip, width=0.5\textwidth]{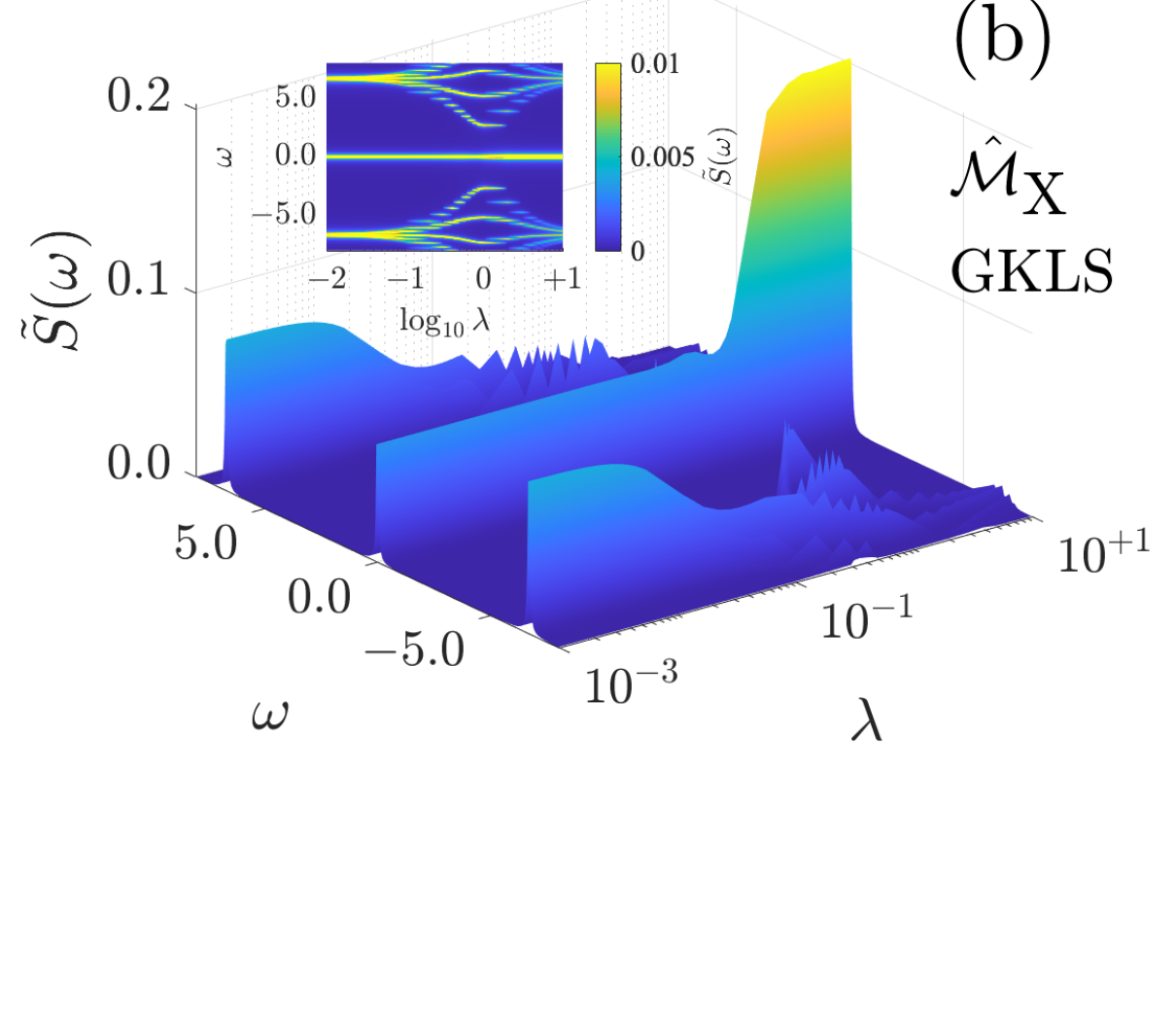}\\
\includegraphics[trim={0.25cm 3cm 0.25cm 9cm},clip, width=0.5\textwidth]{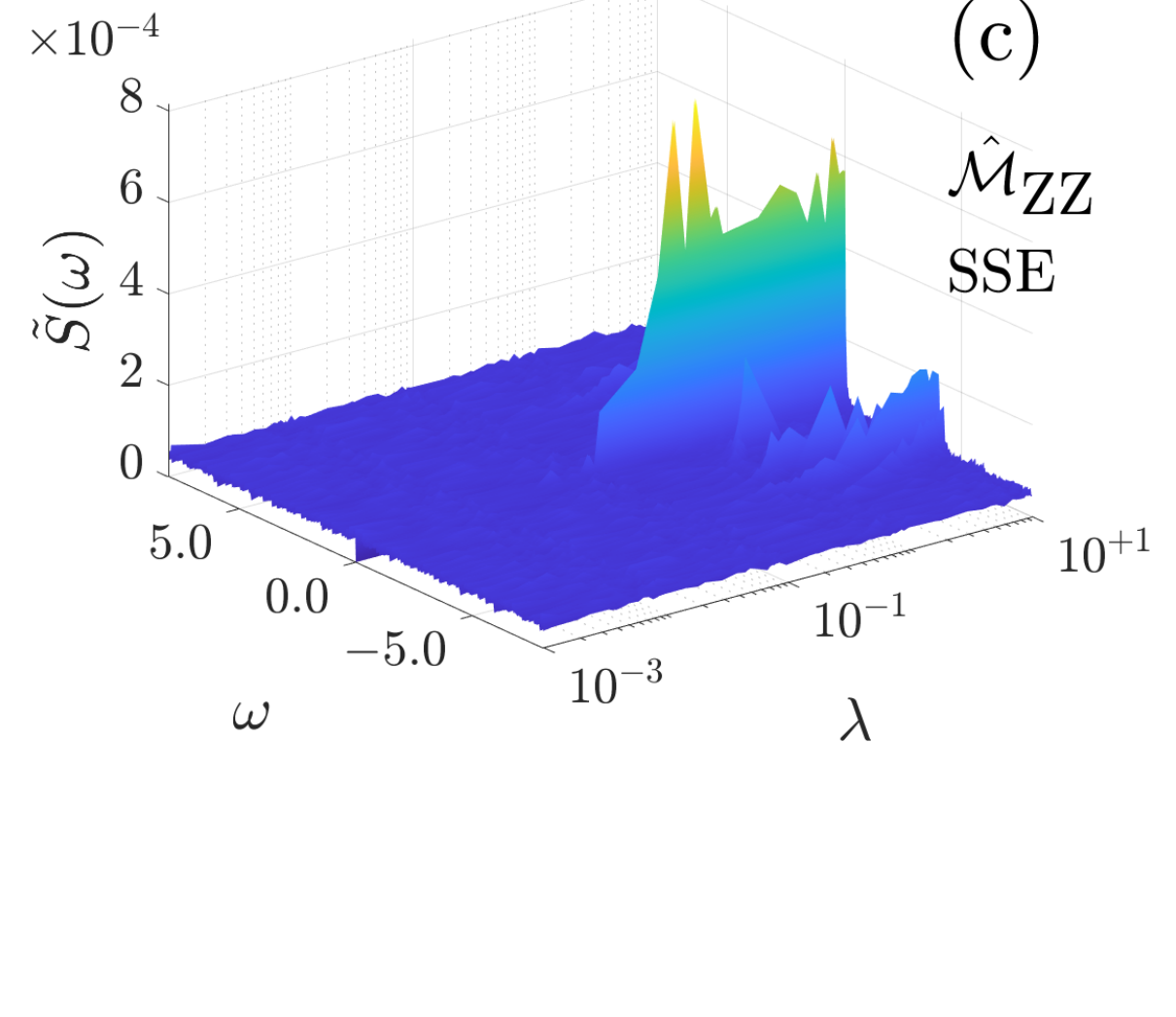}\hfill
\includegraphics[trim={0.25cm 3cm 0.25cm 9cm},clip, width=0.5\textwidth]{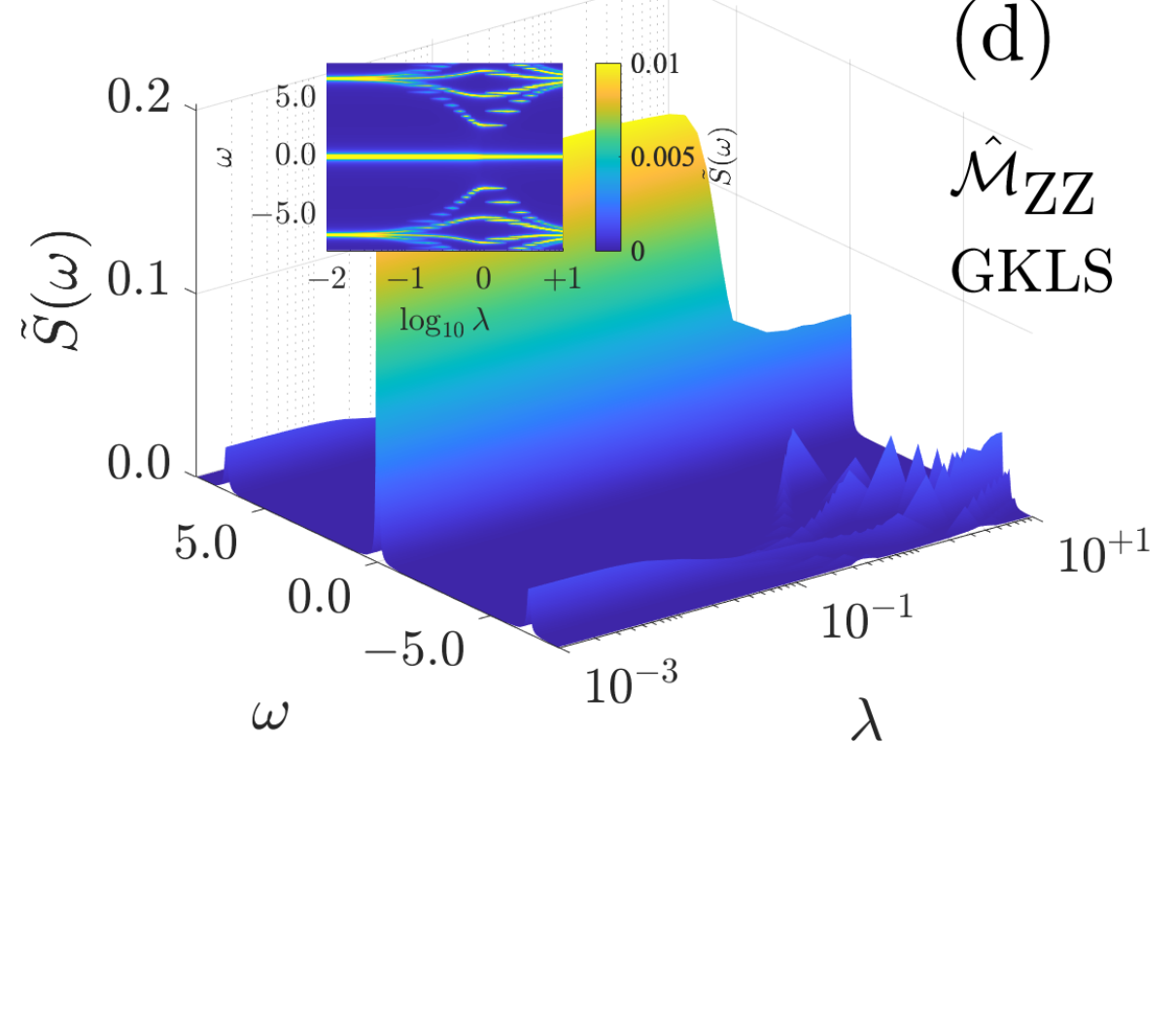}
\caption{We plot normalized PSDs $\tilde{S}(\omega)$ for the transverse-field Ising model, where we measure the observables: $\hat{\mathcal{M}}_\text{X}=\sum_{i=1}^N \sigma_i^x$ and $\hat{\mathcal{M}}_\text{ZZ}=\sum_{i=1}^N \sigma_i^z \sigma_{i+1}^z$. As in Fig.\ \ref{BHMeasureFig}, the spectra in (\textbf{a}) and (\textbf{c}) are obtained by simulating the SSE, whereas (\textbf{b}) and (\textbf{d}) are obtained with Eq.\ \eqref{eq:Sexpression}.  The PSDs for the transverse-field Ising model were obtained for $N = 10$ spins for the SSE calculations and $N = 6$ spins for the ones with Eq.\ \eqref{eq:Sexpression}.  For all these four PSDs, we kept $\gamma = 0.01$.  The insets in (\textbf{b}) and (\textbf{d}) are rotated versions of the main plots. The abrupt change -- cf.\ the order parameter vs $\lambda$ plot in Fig.\ \ref{FigBHIsingOP}(b) -- in the PSDs near $10^{-2} < \lambda < 10^{-1}$ is clearly visible. The height of the peak at $\omega=0$, however, changes continuously near $10^{-1} < \lambda < 1$.}\label{IsingMeasureFig}
\end{figure*}

We now show that the change in the PSD reveals the phase transition in the transverse-field Ising chain, which is exactly solvable using the Jordan-Wigner transformation and is a paradigm for quantum phase transitions \cite{sachdev}. This model was implemented with trapped ions \cite{lanyon2011universal,smith2016many}, where a dynamical phase transition was observed \cite{dynamical}. The Hamiltonian is
\begin{equation}
\hat{\mathcal{H}}=-\sum_{i=1}^N \sigma_i^z \sigma_{i+1}^z - \lambda \sum_{i=1}^N \sigma_i^x,
\label{IsingHMainText}
\end{equation}
where $\sigma_i^{x,z}$ are Pauli operators, we use periodic boundary conditions, and $\lambda$ is a dimensionless parameter. As $\lambda$ is varied, the system exhibits a quantum phase transition at $\lambda_c=1$ in the thermodynamic limit from a ferromagnetic $\lambda<\lambda_c$ to a paramagnetic $\lambda>\lambda_c$ phase. We consider a homodyne measurement of the coupling $\hat{\mathcal{M}}_\text{ZZ}=\sum_{i=1}^N \sigma_i^z \sigma_{i+1}^z$ and transverse-field $\hat{\mathcal{M}}_\text{X}=\sum_{i=1}^N \sigma_i^x$.

The PSDs obtained by numerically integrating the SSE (\ref{eq:homodyne})  for a system with $N=10$ and measurement operators $\hat{\mathcal{M}}_\text{X}$ and $\hat{\mathcal{M}}_\text{ZZ}$ are shown in Fig.\ \ref{IsingMeasureFig}(a,c), respectively.  We also obtain the PSDs using Eq.\ \eqref{eq:Sexpression} for a system with $N=6$ spins for the same measurement operators in Fig.\ \ref{IsingMeasureFig}(b,d).  In order to compare the PSDs for different values of $\lambda$, we always rescale the Hamiltonian such that its spectrum spans the same frequency range (20 in dimensionless units).

Similar to the Bose-Hubbard PSDs, the qualitative nature of the PSDs change when we go from the ferromagnetic to the paramagnetic phase. We note that the $\hat{\mathcal{M}}_\text{X}$ PSDs in the ferromagnetic part is qualitatively similar to the paramagnetic part of the $\hat{\mathcal{M}}_\text{ZZ}$ PSDs.  This is true for all the PSDs.  The ferromagnetic part of $\hat{\mathcal{M}}_\text{ZZ}$ PSD obtained using Eq.\ \eqref{eq:Sexpression} in Fig.\ \ref{IsingMeasureFig}(d) is similar to the paramagnetic $\hat{\mathcal{M}}_\text{X}$ PSDs.

To obtain the PSDs using the SSE (\ref{eq:homodyne}), we start with the ground state of the Hamiltonian (\ref{eq:HamiltonAlpha}) at $t = 0$. Moreover, if $[\hat{\mathcal{H}},\hat{\mathcal{M}}_i] = 0$ -- e.g., when $\lambda = 0$  $(1/\lambda = 0)$ in the $\hat{\mathcal{M}}_\text{ZZ}$  $(\hat{\mathcal{M}}_\text{X})$ measurement in the transverse-field Ising model -- the measurement process does not change the initial wavefunction. This leads to a flat PSD with no features. This is unlike the GKLS PSDs assuming $\rho^{\text{st}} = \mathbb{1}/N$, where $[\hat{\mathcal{H}},\hat{\mathcal{M}}_i] = 0$ results in $S(\omega) \propto \delta(\omega)$.

We have $[\hat{\mathcal{H}},\hat{\mathcal{M}}_i] \neq 0$ for the parameter ranges considered in Figs.\ \ref{BHMeasureFig} and \ref{IsingMeasureFig}. Therefore, the measurement process is equivalent to an exploration of the phase space even if we start with an eigenstate of $\hat{\mathcal{H}}$. However, we believe that the integrability of the transverse-field Ising model is responsible for the absence of any peaks in the ferromagnetic $\hat{\mathcal{M}}_\text{ZZ}$ PSD Fig.\ \ref{IsingMeasureFig}(c) obtained using Eq.\ \eqref{eq:homodyne}. Since we start with a mixed state $\propto \mathbb{1}$ while using Eq.\ \eqref{eq:Sexpression}, the $\hat{\mathcal{M}}_\text{ZZ}$ PSD still has a peak even in the ferromagnetic phase.

\section{Change in PSD  due to the Commutation Relation $[ \hat{\mathcal{H}}, \hat{\mathcal{M}}_0 ]$}

\begin{figure*}
\includegraphics[trim={0.25cm 7cm 0.25cm 7cm},clip, width=0.5\textwidth]{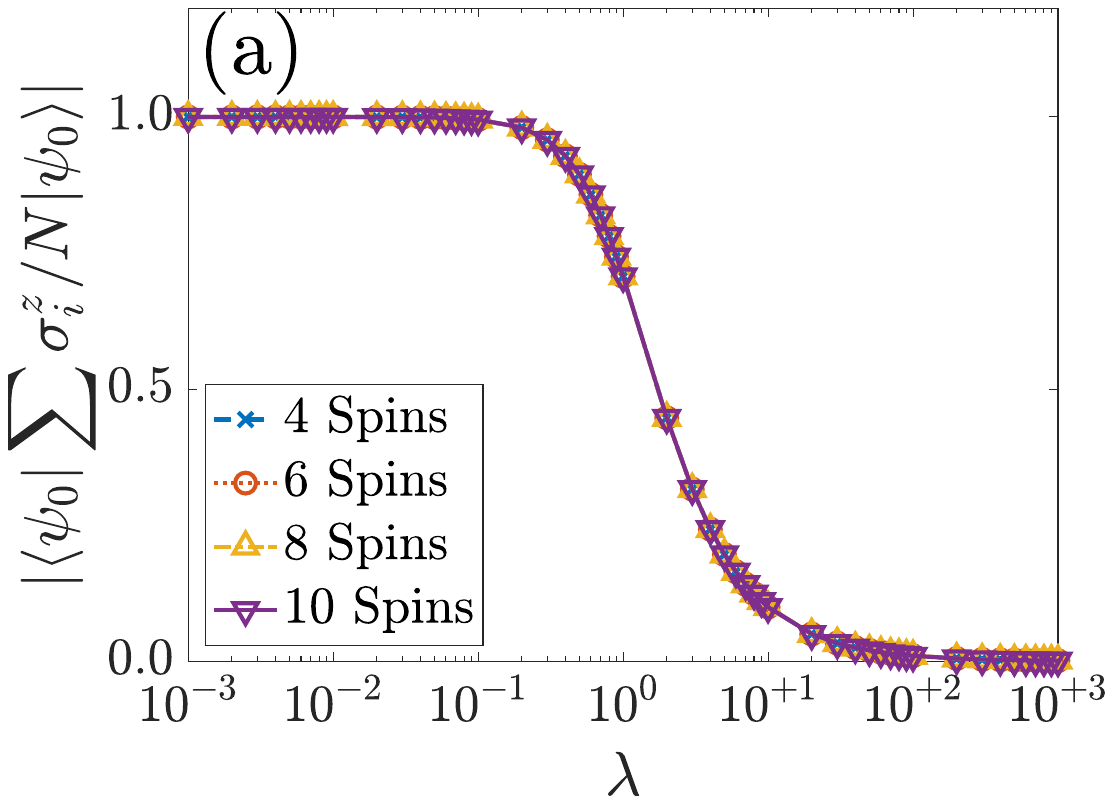}\hfill
\includegraphics[trim={0.25cm 7cm 0.25cm 7cm},clip, width=0.5\textwidth]{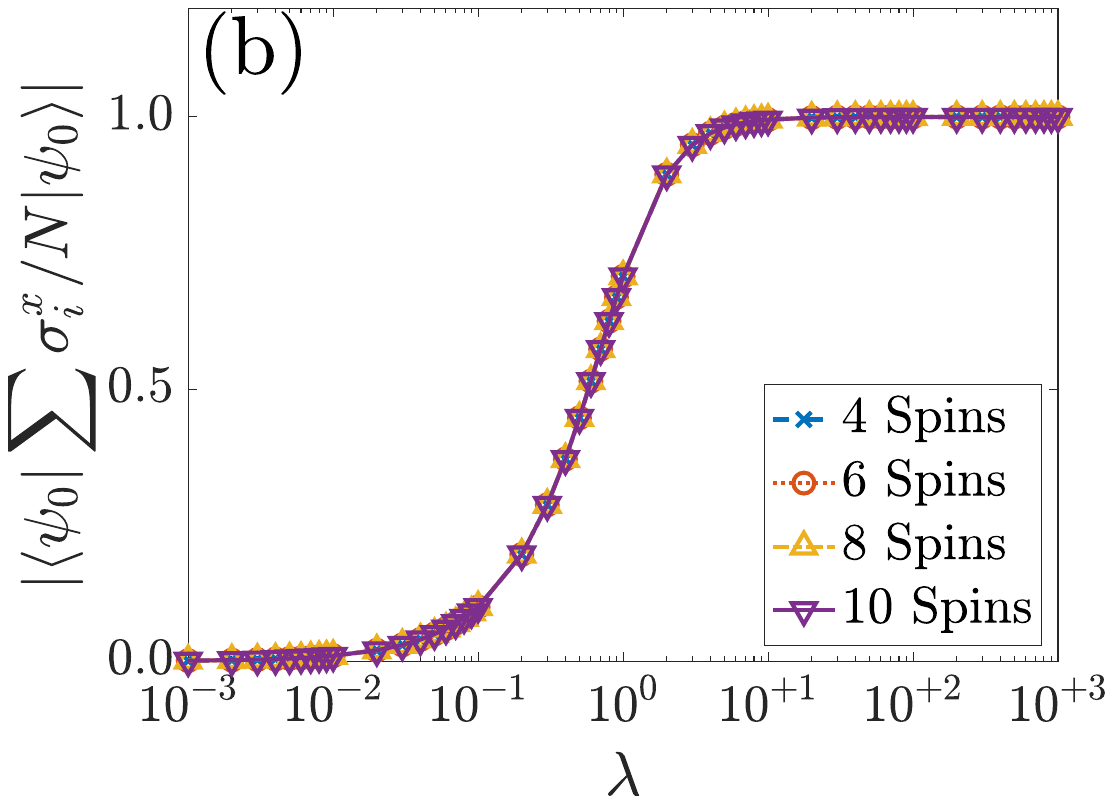}
\caption{We plot $|\langle \psi_0 | \sum_{i}\sigma_{i}^{z}/N | \psi_0 \rangle|$ and $|\langle \psi_0 | \sum_{i}\sigma_{i}^{x}/N | \psi_0 \rangle|$ vs the parameter $\lambda$ for the Hamiltonian \eqref{NPTHMainText} that does not have a phase transition in Figs.\ (\textbf{a}) and (\textbf{b}), respectively.  Here $| \psi_0 \rangle$ is the ground state of the Hamiltonian.  Unlike the abrupt transition (albeit in the log scale) in the order parameter for the transverse-field Ising model in Fig.\ \ref{FigBHIsingOP}(b), here we observe a continuous change. Moreover, as we change the system size, the plots do not change as much as they did for the order parameter in Fig.\ \ref{FigBHIsingOP}(b). In fact,  the behavior of these expectation values is more akin to the inset of Fig.\ \ref{FigBHIsingOP}(b).}
\label{FigNPTOP}
\end{figure*}

\begin{figure*}
\includegraphics[trim={0.25cm 7.5cm 0.25cm 7cm},clip, width=0.5\textwidth]{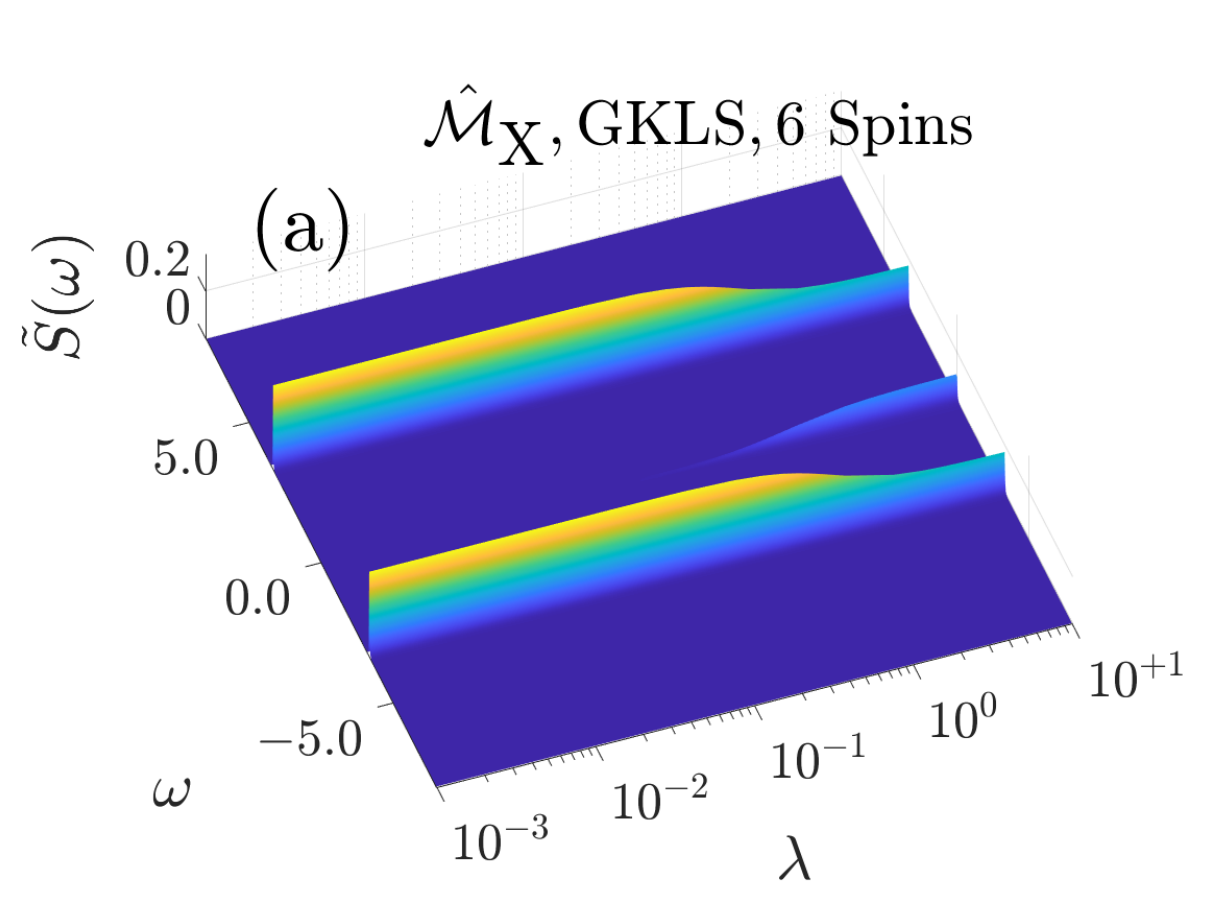}\hfill
\includegraphics[trim={0.25cm 7.5cm 0.25cm 7cm},clip, width=0.5\textwidth]{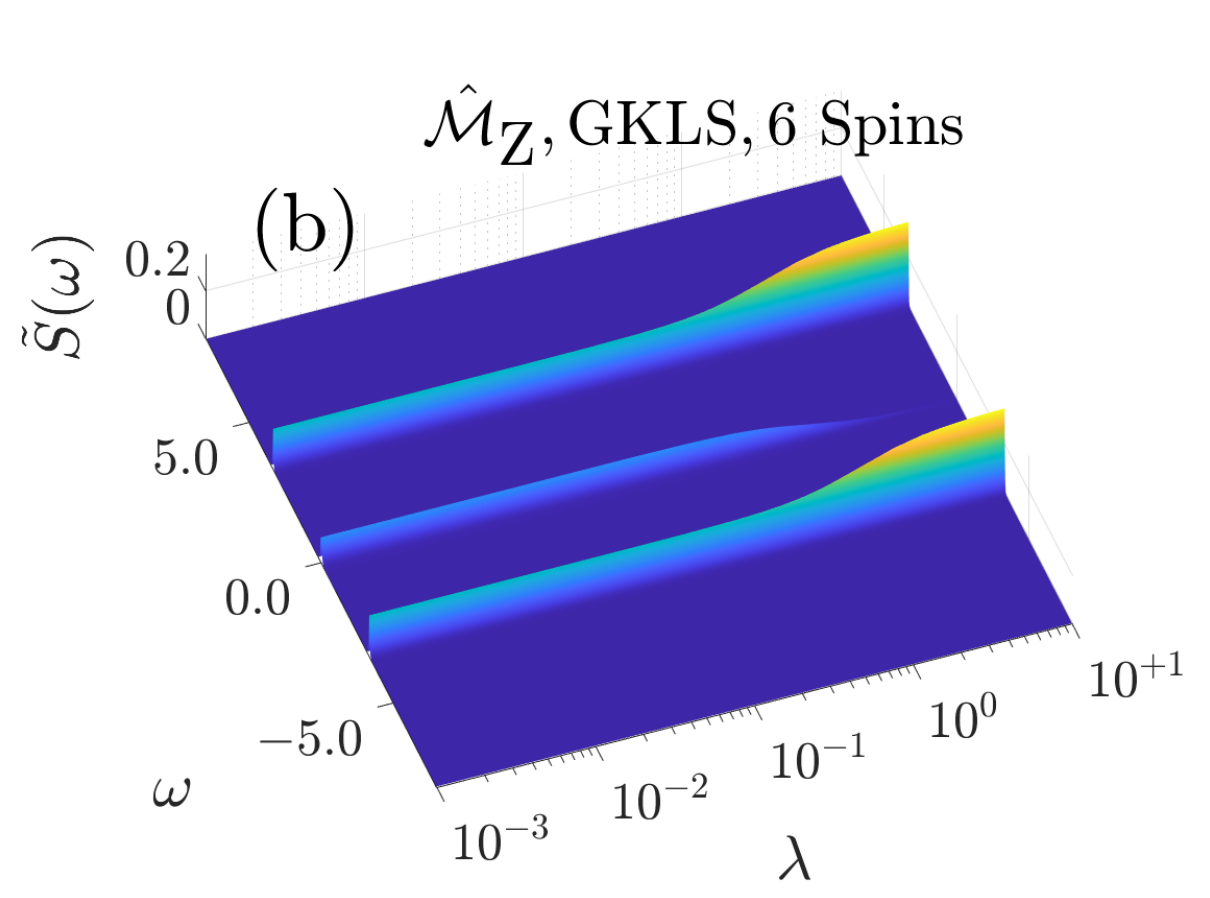}\\
\includegraphics[trim={0.25cm 7.5cm 0.25cm 7cm},clip, width=0.5\textwidth]{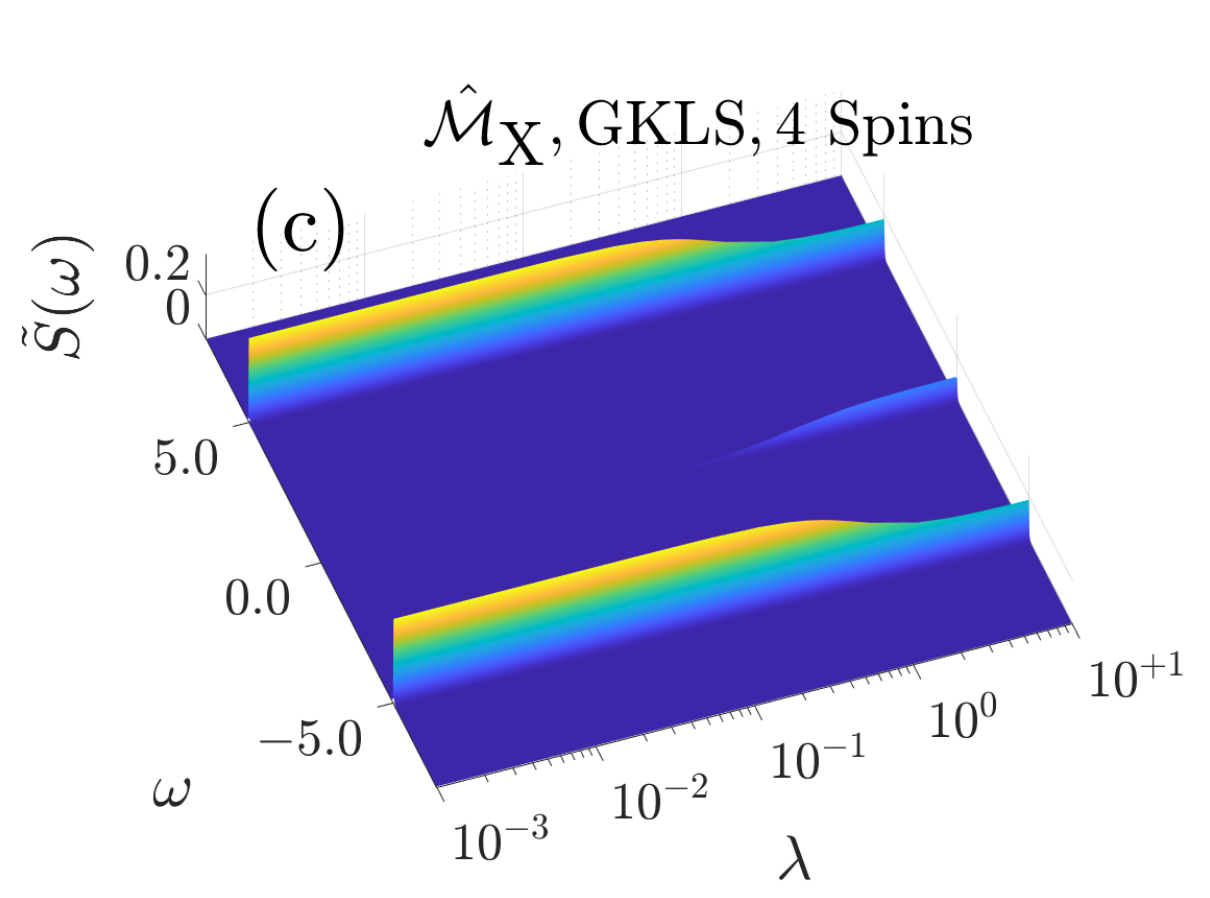}\hfill
\includegraphics[trim={0.25cm 7.5cm 0.25cm 7cm},clip, width=0.5\textwidth]{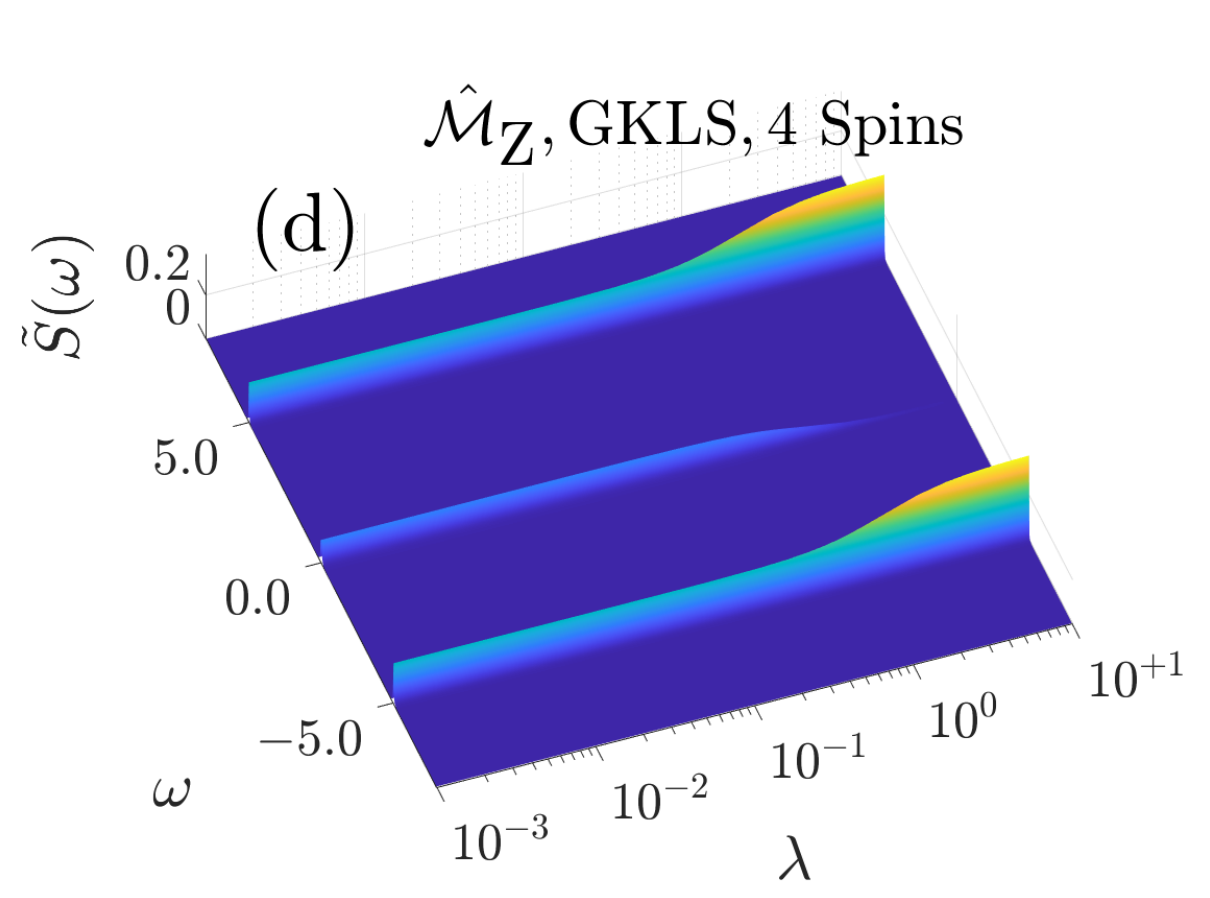}
\caption{We plot normalized PSDs $\tilde{S}(\omega)$ for the Hamiltonian \eqref{NPTHMainText} with no phase transitions, where we measure the observables: $\hat{\mathcal{M}}_\text{X}=\sum_{i=1}^N \sigma_i^x$ and $\hat{\mathcal{M}}_\text{Z}=\sum_{i=1}^N \sigma_i^z$.  In panels (\textbf{a}) and (\textbf{b}), we consider six spins, whereas in panels (\textbf{c}) and (\textbf{d}) we consider four spins.  All the PSDs are obtained with Eq.\ \eqref{eq:Sexpression}.  Similar to Figs.\ \ref{IsingMeasureFig}(b,d), the heights of the three peaks change continuously near $10^{-1} < \lambda < 1$. }\label{FigNPTPSD}
\end{figure*}

In the foregoing analysis, we considered a Hamiltonian that depended on a single parameter $\alpha$.  The measurement operator $\hat{\mathcal{M}}_0$ is chosen such that it commutes with one part of the Hamiltonian $\hat{\mathcal{H}_0}$ while not commuting with the other: $\hat{\mathcal{H}_1}$. Writing the commutation relation between the Hamiltonian and the measurement operator as
\begin{equation}
\left[ \hat{\mathcal{H}}, \hat{\mathcal{M}}_0 \right] = \left[ \hat{\mathcal{H}_0}, \hat{\mathcal{M}}_0 \right] + \alpha \left[ \hat{\mathcal{H}_1}, \hat{\mathcal{M}}_0 \right],
\label{CommHM0}
\end{equation}
we observe that $\hat{\mathcal{M}}_0$ evolves from commuting with $\hat{\mathcal{H}}$ to not commuting as we change $\alpha$.  However, it is important to note that $[ \hat{\mathcal{H}}, \hat{\mathcal{M}}_0 ] \neq 0$ for the range of $\alpha$ considered in Sec.\ \ref{SecBHMainText} with the identification $\alpha \equiv U/J$ and in Sec.\ \ref{SecTFIMainText} with $\alpha \equiv \lambda$.

Nevertheless, one needs to be careful while discerning the changes in PSDs due to phase transitions since in the finite sized systems one expects to see some changes in the PSDs simply because of the commutation properties (e.g., $[ \hat{\mathcal{H}}, \hat{\mathcal{M}}_0 ]$ being equal or unequal to zero). To illustrate this, we consider the Hamiltonian
\begin{equation}
\hat{\mathcal{H}}=-\sum_{i=1}^N \sigma_i^z - \lambda \sum_{i=1}^N \sigma_i^x,
\label{NPTHMainText}
\end{equation}
which does not go through a phase transition.  The ground state of $\hat{\mathcal{H}_0} = -\sum_{i=1}^N \sigma_i^z$ is connected to the ground state of $\hat{\mathcal{H}_1} = -\sum_{i=1}^N \sigma_i^x$ by continuous rotations. This is demonstrated by the ground state expectation values of $\sum_{i}\sigma_{i}^{z}/N$ and $\sum_{i}\sigma_{i}^{x}/N$ in Figs.\ \ref{FigNPTOP}(a) and \ref{FigNPTOP}(b), both of which are similar to the inset of Fig.\ \ref{FigBHIsingOP}(b).

Similar to Sec.\ \ref{SecTFIMainText}, we show the PSDs obtained for the measurements $\hat{\mathcal{M}}_\text{Z}=\sum_{i=1}^N \sigma_i^z$ and $\hat{\mathcal{M}}_\text{X}=\sum_{i=1}^N \sigma_i^x$ in Fig.\ \ref{FigNPTPSD} for the Hamiltonian \eqref{NPTHMainText} by using Eq.\ \eqref{eq:Sexpression}. Here we see a continuous change in the PSDs between $\lambda \approx 0.1$ and $\lambda \approx 1$. Interestingly, this change appears at the same interval in $\lambda$ for six as well as for four spin PSDs. The position of the peaks in $\omega$ are slightly different in the four and the six spin PSDs, whereas the heights remain almost unchanged.

We note that the changes in the PSDs due to the commutation relations are not as abrupt as the ones caused by the change in the Hamiltonian spectrum due to a phase transition.  We believe that the commutation relations change the PSDs trivially compared to the changes occurring due to a phase transition, and these two types of changes in the PSDs can indeed be differentiated.  However, to confirm this hypothesis peremptorily, one either needs to independently verify with an experiment or to perform numerics on a thermodynamically large system.

\section{Summary and Outlook}

We show that it is possible to detect phase transitions in the 1D Bose Hubbard model and the transverse-field Ising model by discerning the qualitative changes in the measurement signals of weak continuous measurements. To observe these changes, one need not prepare the state in a particular way or be confined to the ground state.  We believe that this method of detecting the phase transition can be applied to various strongly interacting systems for a range of experimentally realizable measurement operators.

We have focused on the situation where the system Hamiltonian is known.  In other situations of interest, this might not be the case. It will be interesting to investigate what can be deduced about a system's Hamiltonian from measurement records. Furthermore, our criterion may be generalized to topological \cite{esslinger} and dynamical phase transitions \cite{dynamical}, which have also been implemented successfully.  Further exploration and, in particular, experiments will be needed to assess the broader applicability of continuous measurements as a probe of phase transitions.

\begin{acknowledgments}

This work was supported by the Villum Foundation, the Independent Research Fund Denmark under Grant Number 8049-00074B, the Carlsberg Foundation, and ERC, H2020 grant 639560 (MECTRL). LFB would like to thank the Max Planck Institute for the Physics of Complex Systems for hospitality during visits to the institute.

\end{acknowledgments}

\appendix

\section*{Organization of the Appendices}
\label{SecOrgAppendix}

We start by reviewing a few important properties of the Liouvillian and vectorization (the latter is also known as the Choi-Jami\l kowski isomorphism \cite{suppChoi, suppJam}),  which is used extensively in the following calculations.  We then write the stochastic master equation keeping terms up to order $\sqrt{\text{d} t}$. In the process, we compare the notations of Refs.\ \cite{wiseman2009quantum},  \cite{klaus2}, \cite{jacobsIntro} and \cite{jacobsBook}.  We go over the derivation for the autocorrelation function of the measurement record $F^{(1)}_\textrm{hom}(t,t+\tau)$.  Starting from the expression for $F^{(1)}_\textrm{hom}$ and making use of the quantum regression theorem, we  derive the expression for the PSD -- Eq.\ \eqref{eq:Sexpression} in the main text.  We further simplify this using the Choi-Jami\l kowski isomorphism and obtain Eqs.\ \eqref{eq:eigenPSD} and \eqref{eq:eigenPSDzero} of the main text.  We describe the numerical procedure for obtaining the PSDs in Figs.\ \ref{BHMeasureFig}(b,d), \ref{IsingMeasureFig}(b,d) and \ref{fig:Zeno} in the main text.  We provide an expression and examples of pictorial representations of the matrix elements $M_{jk}$ necessary for constituting the measurement operator $\hat{\mathcal{M}}_0$.  Finally, we describe the numerical integration procedure for obtaining the PSDs in Figs.\ \ref{BHMeasureFig}(a,c) and \ref{IsingMeasureFig}(a,c).

\section{Liouvillian and Vectorization}
\label{SecSuppLiouVec}

Markovian dynamics of a linear and completely positive open quantum system can be described by a Gorini-Kossakowski-Lindblad-Sudarshan (GKLS) master equation \cite{suppGKS, suppLindblad, suppBreuer, wiseman2009quantum, jacobsBook, jacobsIntro}
\begin{equation}
\dot \rho  = - i[\hat{\mathcal{H}},\rho] + \sum_i \gamma_i \mathcal{D}[\hat L_i] \rho(t),
\label{eq:Lindbladian}
\end{equation}
where
\begin{equation}
\mathcal{D}[\hat{L}]\rho = \hat{L}\rho\hat{L}^\dagger - \frac{1}{2} \biggl(\hat{L}^\dagger \hat{L}\rho+\rho\hat{L}^\dagger \hat{L}\biggr).
\label{eq:LindbladSuperOp}
\end{equation}
The GKLS master equation is linear in $\rho$, which allows us to associate it with the so-called Liouvillian superoperator $\mathcal{L}$ satisfying $\partial_t \rho=\mathcal{L} \rho$. The superoperator $\mathcal{L}$ is trace preserving and generates the following completely positive trace preserving map $e^{\mathcal{L}t}$ describing the time evolution of the system:
\begin{equation}
\rho(t)  = e^{\mathcal{L}t}\rho(0) = \sum_{i}\hat{K}_{i}(t)\rho(0)\hat{K}^{\dag}_{i}(t),
\label{eq:OpSumRep}
\end{equation}
such that
\begin{equation}
\sum_{i}\hat{K}^{\dag}_{i}(t)\hat{K}_{i}(t) =  \mathbb{1},
\label{eq:OpSumRepIdentity}
\end{equation}
where the set of operators $\lbrace \hat{K}_{i} \rbrace$ are called Kraus operators.  The above way of representing the completely positive trace preserving map is called the operator-sum representation.

Superoperators such as $\mathcal{L}$ act on the Liouville space $\mathcal{B}(\mathcal{H})$ consisting of all the linear operators acting on the Hilbert space. This space can itself be treated as a Hilbert space with the Hilbert-Schmidt inner-product $\langle \langle \hat{A} | \hat{B} \rangle \rangle=\text{Tr}(\hat{A}^\dagger \hat{B})$. We use the notation $|\rho\rangle \rangle$ for a vectorized state that is created by stacking the columns of $\rho$. In order to ease the calculations, we apply this vectorized notation here \cite{suppVector}. The vectorized representation of $\mathcal{L}$ is
\begin{multline}
\mathbb{L} = -i (\mathbb{1}\otimes\hat{\mathcal{H}} - \hat{\mathcal{H}}^T\otimes \mathbb{1}) \\
+\sum_i\frac{\gamma_i}{2}\left(2 \hat{L}_i^*\otimes\hat{L}_i  - \mathbb{1} \otimes \hat{L}_i^\dagger \hat{L}_i - \hat{L}_i^T \hat{L}_i^* \otimes \mathbb{1} \right), \label{eq:matrixRep}
\end{multline}
where $A^T$ denotes the transpose.  Note that generally $\mathbb{L}$ is a non-Hermitian matrix.

In this paper, we are only concerned with diagonalizable Liouvillians.  For non-diagonalizable Liouvillians, one needs to consider the Jordan normal form. Unlike the Hamiltonian, the Liouvillian is generally not Hermitian, i.e., the adjoint superoperator $\mathcal{L}^\dagger$ is not equal to $\mathcal{L}$. For this reason, the eigenvalues of $\mathcal{L}$ are generally complex and it has different right and left eigenstates satisfying
\begin{subequations}
\begin{align}
\mathbb{L} |r_m\rangle \rangle &= \lambda_m |r_m\rangle \rangle, \label{eq:right} \\
\mathbb{L}^\dagger |l_m\rangle \rangle &= \lambda_m^* |l_m\rangle \rangle. \label{eq:left}
\end{align}
\end{subequations}
We fix the normalization such that the left and right eigenstates are orthonormal $\langle \langle r_m| l_n \rangle \rangle = \delta_{mn}$, which is called the biorthogonality.  Enumerating the eigenstates according to Eqs.\ \eqref{eq:right} and \eqref{eq:left}, we obtain the following completeness relation:
\begin{equation}
\sum_m |r_m\rangle \rangle \langle \langle l_m| = \mathbb{1}. \label{eq:unity}
\end{equation}

We assume that the open system dynamics are due to a continuous weak measurement of a single Hermitian operator $\hat{L} = \hat{\mathcal{M}}_0$. At long times, the unmonitored system will reach a steady state of $\mathcal{L}$ defined by $\mathcal{L}[\rho^{\text{st}}]=0$, i.e.,  a member of the kernel of the operator  $\mathcal{L}$.  For a Hermitian operator, we make the simple observation that
\begin{multline}
\mathcal{L}[\mathbb{1}] = - [\hat{\mathcal{H}},\mathbb{1}] +\gamma\hat{\mathcal{M}}_0 \mathbb{1} \hat{\mathcal{M}}_0 \\
- \frac{\gamma}{2}\biggl(\hat{\mathcal{M}}_0^2 \mathbb{1} + \mathbb{1} \hat{\mathcal{M}}^2_0\biggr) = 0,
\end{multline}
which shows that $\mathbb{1}/N$ is always a stationary state where \textit{N} is the dimension of the Hilbert space. In general, there is no guarantee that this is the only stationary state, but we assume
\begin{equation}
\rho^{\text{st}} = \mathbb{1}/N
\label{SSDef}
\end{equation}
for simplicity.

\section{Stochastic Master Equation}
\label{SecSME}

We write the stochastic master equation (SME) corresponding to the stochastic Schr\"{o}dinger equation (SSE) considered in the main text.  In the SME, we only keep the terms upto order $\sqrt{dt}$.  In the process, we reconcile the derivations and notations of Refs.\ \cite{wiseman2009quantum}, \cite{jacobsBook}, \cite{jacobsIntro} and \cite{klaus2}.

The definition of homodyne current in Refs.\ \cite{klaus2}, \cite{jacobsIntro} and \cite{jacobsBook} is as follows:
\begin{equation}
\lambda_{t}[\hat{O}] = \left\langle \hat{O} \right\rangle_{\rho} \text{d} t + \frac{\text{d} W_{t}}{\sqrt{8\mathcal{K}}}.
\label{Hom_Curr_Blatt}
\end{equation}
The corresponding SSE is
\begin{multline}
\text{d} \left | \bar{\psi}\left(t\right) \right \rangle  = \left\lbrace - iH\text{d} t -\mathcal{K} \hat{O}^{2}\text{d} t \right. \\
\left. + 4\mathcal{K}\hat{O}\lambda_{t}[\hat{O}]  \right\rbrace \left | \bar{\psi}\left(t\right) \right \rangle,
\label{SSE_Blatt}
\end{multline}
where  $\left | \bar{\psi}\left(t\right) \right \rangle$ symbolizes the non-normalized wavefunction. We obtain the homodyne current and the SSE of the main text from Eqs.\ \eqref{Hom_Curr_Blatt} and (\ref{SSE_Blatt}) as follows:
\begin{subequations}
\begin{align}
\textrm{define: }& I(t)\text{d} t = \frac{\gamma}{2}\lambda_{t}[\hat{O}]; \\
\textrm{replace: }& \hat{O} \rightarrow 4\hat{\mathcal{M}}_{0},\; \mathcal{K} \rightarrow \frac{\gamma}{32}.
\label{Pres}
\end{align}
\end{subequations}
While applying the prescription (\ref{Pres}) in Eq.\ \eqref{SSE_Blatt}, one does not change the wavefunction $ \left | \bar{\psi}\left(t\right) \right \rangle$.

We now apply the prescription (\ref{Pres}) to the SME of Ref.\ \cite{klaus2}.  Before we do so, we clarify the different definitions of the Lindblad superoperator appearing in different references. We list all the definitions below as
\begin{widetext}
\begin{subequations}
\begin{align}
\textrm{Refs.\ \cite{wiseman2009quantum}, \cite{jacobsBook}, \cite{jacobsIntro} and the current manuscript:}\quad & \mathcal{D}[\hat{O}]\rho = \hat{O}\rho \hat{O}^{\dagger} -\frac{1}{2}\left\lbrace \hat{O}^{\dagger}\hat{O}, \rho\right\rbrace,\label{Drhojs} \\
\textrm{Ref.\ \cite{klaus2}:}\quad & \mathcal{D}[\hat{O}]\rho = 2\hat{O}\rho \hat{O}^{\dagger} -\left\lbrace \hat{O}^{\dagger}\hat{O}, \rho\right\rbrace.\label{DrhoBlatt}
\end{align}
\label{Drho}
\end{subequations}
\!\!For the rest of the discussion, we will be using the definition (\ref{Drhojs}).  The SME conditioned on the random measurement outcome (\ref{Hom_Curr_Blatt}) is
\begin{equation}
\text{d} \rho = -\frac{i}{\hbar}[H, \rho]\;\text{d} t + 2\mathcal{K} \mathcal{D}[\hat{O}] \rho \; \text{d} t +\underbrace{4\mathcal{K}\mathcal{H}[\hat{O}]\rho  \left(\lambda_{t}[\hat{O}] - \left\langle \hat{O} \right\rangle_\rho \text{d} t \right)}_{= \sqrt{2\mathcal{K}}\mathcal{H}[\hat{O}]\rho\text{d} W_{t}},
\label{SMEBlatt}
\end{equation}
where $\mathcal{H}[\hat{O}]\rho =  \hat{O}\rho +  \rho\hat{O}^{\dagger} - \left\langle \hat{O} + \hat{O}^{\dagger}\right\rangle_{\rho}\rho$.  All the references agree on the definition of $\mathcal{H}[\hat{O}]\rho$. Here, we consider the detector to be $100\%$ efficient. Also, since we are using the definition (\ref{Drhojs}), the coefficient of the second term is $2\mathcal{K}$ instead of $\mathcal{K}$ (cf.  Eq.\ (5) of \cite{klaus2}).

Using the prescription (\ref{Pres}) we replace $\sqrt{2\mathcal{K}}$ by $\sqrt{\gamma}/4$ and $\hat{O}$ by $4\hat{\mathcal{M}}_{0}$.  Additionally, we write $\left\langle \cdots \right\rangle_{\rho}$ as $\left\langle \cdots \right\rangle$ for notational convenience. Finally, we obtain
\begin{equation}
\begin{split}
\rho(t + \text{d} t) &= \rho(t) + \left(-\frac{i}{\hbar}[H, \rho] + \gamma\mathcal{D}[\hat{\mathcal{M}}_0]\rho\right)\text{d} t + \sqrt{\gamma}\mathcal{H}[\hat{\mathcal{M}}_{0}]\rho\; \text{d} W_{t} \\
&\approx \rho(t) +  \sqrt{\gamma}\left( \hat{\mathcal{M}}_{0}\rho(t) + \rho(t)\hat{\mathcal{M}}^{\dagger}_{0}\right) \text{d} W_{t} - \sqrt{\gamma}\left\langle\hat{\mathcal{M}}_{0} + \hat{\mathcal{M}}^{\dagger}_{0}\right\rangle\text{d} W_{t},
\end{split}
\label{ModSME}
\end{equation}
where we have retained terms only upto order $\sqrt{\text{d} t}$.  Later in this manuscript,  we consider $\rho(t)$ is a priori known to be $\rho^{\text{st}} = \mathbb{1}/N$.  In the above equation,  $\rho(t + \text{d} t)$ is conditioned on the homodyne current until time $t$.

\section{Output Field Correlation Function}
\label{SecF1}

We revisit the derivation for the autocorrelation function of the measurement record $F^{(1)}_\textrm{hom}(t,t+\tau)=\mathbb{E}[I(t+\tau)I(t)]$.  Note that we write the autocorrelation function as $F^{(1)}$. This is because of its relation to Glauber's first-order coherence function. Here we follow Ref.\ \cite{wiseman2009quantum} closely. The steps are as follows:
\begin{equation}\label{Autocorr}
\begin{split}
F^{(1)}_\textrm{hom}(t,t+\tau)(\text{d} t)^2 &= \mathbb{E}\left[I(t+\tau)I(t)\right](\text{d} t)^2 \\
&= \frac{\gamma^2}{4}\mathbb{E}\left[\lambda_{t+\tau}[\hat{O}]\lambda_{t}[\hat{O}]\right] \\
&= \frac{\gamma^2}{4}\mathbb{E}\left[ \left(4\left\langle \hat{\mathcal{M}}_{0} \right\rangle(t+\tau)\;\text{d} t + \frac{\text{d} W_{t+\tau}}{\sqrt{\gamma/4}} \right)\left(4\left\langle \hat{\mathcal{M}}_{0} \right\rangle(t)\;\text{d} t + \frac{\text{d} W_{t}}{\sqrt{\gamma/4}} \right) \right] \\
&= \frac{2\gamma^2}{\sqrt{\gamma}}\mathbb{E}\left[ \left\langle \hat{\mathcal{M}}_{0} \right\rangle(t+\tau)\;\text{d} W_{t} \right]\text{d} t + \underbrace{\gamma \mathbb{E}\left[   \text{d} W_{t+\tau} \; \text{d} W_{t}   \right]}_{= \gamma\delta(\tau)(\text{d} t)^2} + 4\gamma^2\mathbb{E}\left[\left\langle \hat{\mathcal{M}}_{0} \right\rangle(t+\tau)\right]\left\langle \hat{\mathcal{M}}_{0} \right\rangle(t)(\text{d} t)^2.
\end{split}
\end{equation}
While obtaining $\left\langle \hat{\mathcal{M}}_{0} \right\rangle(t+\tau)$,  the trace is calculated with the density operator of Eq.\ \eqref{ModSME} and by identifying $\text{d} t$ with $\tau$. The factorization in the last term of the last line is justified because $\rho(t)$ is given.  Using similar argument we have
\begin{equation}
\mathbb{E}\left[\text{d} W_{t+\tau}\left\langle \hat{\mathcal{M}}_{0} \right\rangle(t)\right] = \underbrace{\mathbb{E}\left[\text{d} W_{t+\tau}\right]}_{=0}\left\langle \hat{\mathcal{M}}_{0} \right\rangle(t) = 0.
\label{4thterm}
\end{equation}

We explicitly calculate the first term of the last line in Eq.\ \eqref{Autocorr} as follows:
\begin{equation}\label{1Autocorr}
\begin{split}
\mathbb{E}\left[ \left\langle \hat{\mathcal{M}}_{0} \right\rangle(t+\tau)\text{ d} W_{t} \right]\text{d} t &= \textrm{Tr}\left[\hat{\mathcal{M}}_{0}e^{\mathcal{L}\tau} \mathbb{E}\left[ \left\lbrace 1 + \sqrt{\gamma}\text{ d} W_{t}\mathcal{H}[\hat{\mathcal{M}}_{0}] \right\rbrace \rho(t) \text{ d} W_{t} \right]\right]\text{d} t \\
&= \sqrt{\gamma}\textrm{ Tr}\left[ \hat{\mathcal{M}}_{0}e^{\mathcal{L}\tau} \left( \hat{\mathcal{M}}_{0}\rho(t) + \rho(t)\hat{\mathcal{M}}_{0}\right)  \right](\text{d} t)^2 - 2\sqrt{\gamma}\textrm{ Tr}\left[\hat{\mathcal{M}}_{0}e^{\mathcal{L}\tau}\rho(t) \right]\left\langle \hat{\mathcal{M}}_{0} \right\rangle(t)(\text{d} t)^2.
\end{split}
\end{equation}
In the first line, $e^{\mathcal{L}\tau}$ provides the noise averaged time evolution between $t+\text{d} t$ and $t+\tau$. In the final line, we used the
It\^{o} rule.  While expanding the superoperator $\mathcal{H}[\hat{\mathcal{M}}_{0}]$, we also used the fact that $\hat{\mathcal{M}}_{0}$ is self-adjoint. Substituting Eq.\ \eqref{1Autocorr} into Eq.\ \eqref{Autocorr}, we obtain
\begin{equation}\label{2Autocorr}
\begin{split}
F^{(1)}_\textrm{hom}(t,t+\tau)(\text{d} t)^2 &= 2\gamma^2    \textrm{ Tr}\left[    \hat{\mathcal{M}}_{0}e^{\mathcal{L}\tau} \left( \hat{\mathcal{M}}_{0}\rho(t) + \rho(t)\hat{\mathcal{M}}_{0}\right)    \right](\text{d} t)^2    -    4\gamma^2  \textrm{ Tr}\left[    \hat{\mathcal{M}}_{0}e^{\mathcal{L}\tau}\rho(t) \right]    \left\langle \hat{\mathcal{M}}_{0} \right\rangle(t)(\text{d} t)^2     \\
&+ \gamma\delta(\tau)(\text{d} t)^2    +    4\gamma^2\mathbb{E}\left[\left\langle \hat{\mathcal{M}}_{0} \right\rangle(t+\tau)\right]\left\langle \hat{\mathcal{M}}_{0} \right\rangle(t)(\text{d} t)^2.
\end{split}
\end{equation}
To simplify the last term of Eq.\ \eqref{2Autocorr}, we note the following:
\begin{equation}\label{3Autocorr}
\begin{split}
\mathbb{E}\left[    \left\langle \hat{\mathcal{M}}_{0} \right\rangle(t+\tau)    \right]\left\langle \hat{\mathcal{M}}_{0} \right\rangle(t) &= \textrm{ Tr}\left[    \hat{\mathcal{M}}_{0}e^{\mathcal{L}\tau} \mathbb{E}\left\lbrace    \left( 1 + \sqrt{\gamma}\text{ d} W_{t}\mathcal{H}[\hat{\mathcal{M}}_{0}] \right) \rho(t)    \right\rbrace    \right]\left\langle \hat{\mathcal{M}}_{0} \right\rangle(t) \\
&=  \textrm{ Tr}\left[    \hat{\mathcal{M}}_{0}e^{\mathcal{L}\tau}  \rho(t)       \right]\left\langle \hat{\mathcal{M}}_{0} \right\rangle(t) +   \sqrt{\gamma}\textrm{ Tr}\left[    \hat{\mathcal{M}}_{0}e^{\mathcal{L}\tau}    \underbrace{\mathbb{E}\left(    \text{d} W_{t}\mathcal{H}[\hat{\mathcal{M}}_{0}]\rho(t)    \right)}_{= 0}    \right]\left\langle \hat{\mathcal{M}}_{0} \right\rangle(t) \\
&= \textrm{Tr}\left[    \hat{\mathcal{M}}_{0}e^{\mathcal{L}\tau}  \rho(t)       \right]\left\langle \hat{\mathcal{M}}_{0} \right\rangle(t),
\end{split}
\end{equation}
where in the second to last line while performing the $\mathbb{E}$ operation,  we recall that $\text{d} W_{t}$ and $\mathcal{H}[\hat{\mathcal{M}}_{0}]\rho(t)$ are statistically independent.  Using Eq.\ \eqref{3Autocorr} into Eq.\ \eqref{2Autocorr} and substituting $\rho(t) = \rho^{\text{st}} = \mathbb{1}/N$, we finally obtain the autocorrelation function as
\begin{equation}\label{Autocorr_Final}
F^{(1)}_\textrm{hom}(t,t+\tau) = \gamma^{2}\textrm{ Tr}\left[ \left( \hat{\mathcal{M}}_{0} + \hat{\mathcal{M}}^{\dagger}_{0} \right) e^{\mathcal{L}\tau} \left( \hat{\mathcal{M}}_{0}\rho^{\text{st}} + \rho^{\text{st}}\hat{\mathcal{M}}^{\dagger}_{0} \right)  \right] + \gamma\delta(\tau) = \frac{4\gamma^{2}}{N}\langle \langle \hat{\mathcal{M}}_0| e^{\mathbb{L}\tau}| \hat{\mathcal{M}}_0\rangle\rangle+ \gamma\delta (\tau),
\end{equation}
where we have used vectorization and that $\hat{\mathcal{M}}_{0}$ is self-adjoint to write the final expression.  The $\delta$-function in this formula arises due to the local oscillator shot noise or vacuum noise.

\section{Homodyne Spectrum: Exact Analytical Result}
\label{SecSpecAnalytical}

The power spectral density is the Fourier transformation of $F^{(1)}_\textrm{hom}(t,t+\tau)$ with the $\delta$-function dropped. First, note that the result of $e^{\mathcal{L}\tau}$ acting on the Hermitian operator $\hat{\mathcal{M}}_{0}\rho^{\text{st}} + \rho^{\text{st}}\hat{\mathcal{M}}^{\dagger}_{0}$ can be written using the operator-sum representation, see Eq.\ \eqref{eq:OpSumRep}.  As a result,  the autocorrelation function (without the $\delta$-function) is the trace of a Hermitian operator, which is real.  Moreover, the autocorrelation function is an even function in $\tau$. Using these properties, we obtain
\begin{equation}
S(\omega) = \frac{8\gamma^{2}}{N} \mathrm{Re} \left[\int_0^\infty \langle \langle \hat{\mathcal{M}}_0| e^{\mathbb{L}\tau}| \hat{\mathcal{M}}_0\rangle\rangle e^{-i\omega \tau} \text{d}\tau \right] = \frac{8\gamma^{2}}{N} h(\omega,\hat{\mathcal{M}}_0,\hat{\mathcal{M}}_0),
\label{eq:Shom}
\end{equation}
where
\begin{equation}
h(\omega,\hat{A},\hat{B}) = \mathrm{Re} \left[\int_0^\infty \langle \langle \hat{A} |e^{\mathbb{L} \tau} |\hat{B}\rangle \rangle e^{-i\omega \tau} \text{d}\tau \right] = \mathrm{Re} \left[\sum_m \langle \langle \hat{A} |r_m\rangle\rangle \langle \langle l_m|\hat{B}\rangle \rangle \int_0^\infty e^{[\mathrm{Re} (\lambda_m) + i(\mathrm{Im} (\lambda_m) - \omega)]\tau}\text{d}\tau \right].
\label{eq:hStep}
\end{equation}
Here we have inserted unity \eqref{eq:unity} and used that $|r_m\rangle\rangle$ is a right eigenstate.

For calculating the integral we must consider two special cases
\begin{subequations}
\begin{align}
&\int_0^\infty e^{[\mathrm{Re} (\lambda_m) + i(\mathrm{Im} (\lambda_m) - \omega)]\tau}\text{d}\tau = \frac{1}{i\omega - \lambda_m} && \text{for} \quad \mathrm{Re}(\lambda_m) <0, \label{eq:firstIntegral} \\
&\int_0^\infty e^{i(\mathrm{Im} (\lambda_m) - \omega)\tau}\text{d}\tau = \pi \delta\left(\omega - \mathrm{Im} (\lambda_m)\right) + \mathcal{P} \left(\frac{1}{i\omega - i\mathrm{Im} (\lambda_m)}\right) && \text{for} \quad \mathrm{Re} (\lambda_m) = 0,
\end{align}
\end{subequations}
where $\mathcal{P}$ denotes the Cauchy principal value. We do not need to consider the case for $\mathrm{Re}(\lambda_m)>0$ since $\mathbb{L}$ cannot have eigenvalues with positive real part. Notice the Cauchy principal value is similar to the result in Eq.~\eqref{eq:firstIntegral} but there is an additional $\delta$-function contribution in the second integral. Plugging this into Eq.~\eqref{eq:hStep} gives the result
\begin{multline}
h(\omega,A,B)=\sum_{\mathrm{Re} (\lambda_m)<0} \frac{-\mathrm{Re} (\lambda_m) \mathrm{Re} (t_m) + \left[\omega - \mathrm{Im} (\lambda_m)\right]\mathrm{Im} (t_m)}{\left[\omega - \mathrm{Im} (\lambda_m)\right]^2 + \left[\mathrm{Re} (\lambda_m)\right]^2} \\
+ \sum_{\mathrm{Re} (\lambda_m) = 0} \left[\pi \mathrm{Re} (t_m)  \delta\left(\omega - \mathrm{Im} (\lambda_m)\right) + \mathcal{P} \left(\frac{\mathrm{Im} (t_m)}{\omega - \mathrm{Im} (\lambda_m)}\right)\right] \label{eq:Hequation}
\end{multline}
where $t_m = \text{Tr}[A^\dagger r_m]\text{Tr}[l_m^\dagger B]$. This is also the result given in the main text in Eqs.\ \eqref{eq:eigenPSD} and \eqref{eq:eigenPSDzero}.
\end{widetext}

\section{Homodyne Spectrum: Numerical Computation}
\label{SecSpecNum}

In this section, we describe how we compute $S(\omega)$ numerically. In principle, we could diagonalize $\mathbb{L}$ and utilize Eqs.\ \eqref{eq:eigenPSD} and \eqref{eq:eigenPSDzero} of the main text. However, this is not feasible due to the large dimensionality of the Liouville space. Instead, from Eq.\ \eqref{eq:hStep} we observe that
\begin{multline}
S(\omega) = \frac{8\gamma^{2}}{N} \mathrm{ Re} \left[\langle \langle \hat{\mathcal{M}_0} |(i\omega \mathbb{1} - \mathbb{L})^{-1} | \hat{\mathcal{M}_0} \rangle \rangle\right] \\
= \frac{8\gamma^{2}}{N}\mathrm{ Re} \left[\langle \langle \hat{\mathcal{M}}_0 | \hat{\xi} \rangle \rangle\right].
\end{multline}
For numerical convenience, we have introduced $|\hat{\xi}\rangle\rangle$ as the solution to the linear equation system
\begin{equation}
(i\omega \mathbb{1} - \mathbb{L}) |\hat{\xi}\rangle\rangle = |\hat{\mathcal{M}}_0\rangle \rangle. \label{eq:calStepBI}
\end{equation}

This equation must be solved for each value of $\omega$. The matrix $(i \omega \mathbb{1} - \mathbb{L})$ preserves the sparsity of the original Hamiltonian.  Even then, we could only compute the numerical spectrum for the 1D Bose-Hubbard model with four sites and four particles, and for the transverse-field Ising model with $N = 6$ spins.  Recall that we could simulate the SSE for the 1D Bose-Hubbard model with six sites and six particles, and for the transverse-field with $N = 10$ spins.

We observe from Eqs.\ \eqref{eq:Shom} and (\ref{eq:Hequation}) that $S(\omega)$ is singular if $\mathrm{Re} (\lambda_m) = 0$ and $\mathrm{Im} (\lambda_m) = \omega$. The system cannot be solved for $\omega=0$, since $\mathbb{L}$ is singular. In the transverse-field Ising (Bose-Hubbard) model, we solve Eq.\ \eqref{eq:calStepBI} for 204 (818) linearly spaced values of $\omega$ between 0.04 (0.01) and 8.00 in dimensionless units. We do not encounter any singularities for these frequency grids. However, in our numerical experience, the system becomes much harder to solve as $\omega \rightarrow 0$.

\section{The Measurement Operators in the Bose-Hubbard Model}
\label{SecImplMeasure}

\begin{figure*}
\includegraphics[trim={0.25cm 7.5cm 0.25cm 7cm},clip, width=0.5\textwidth]{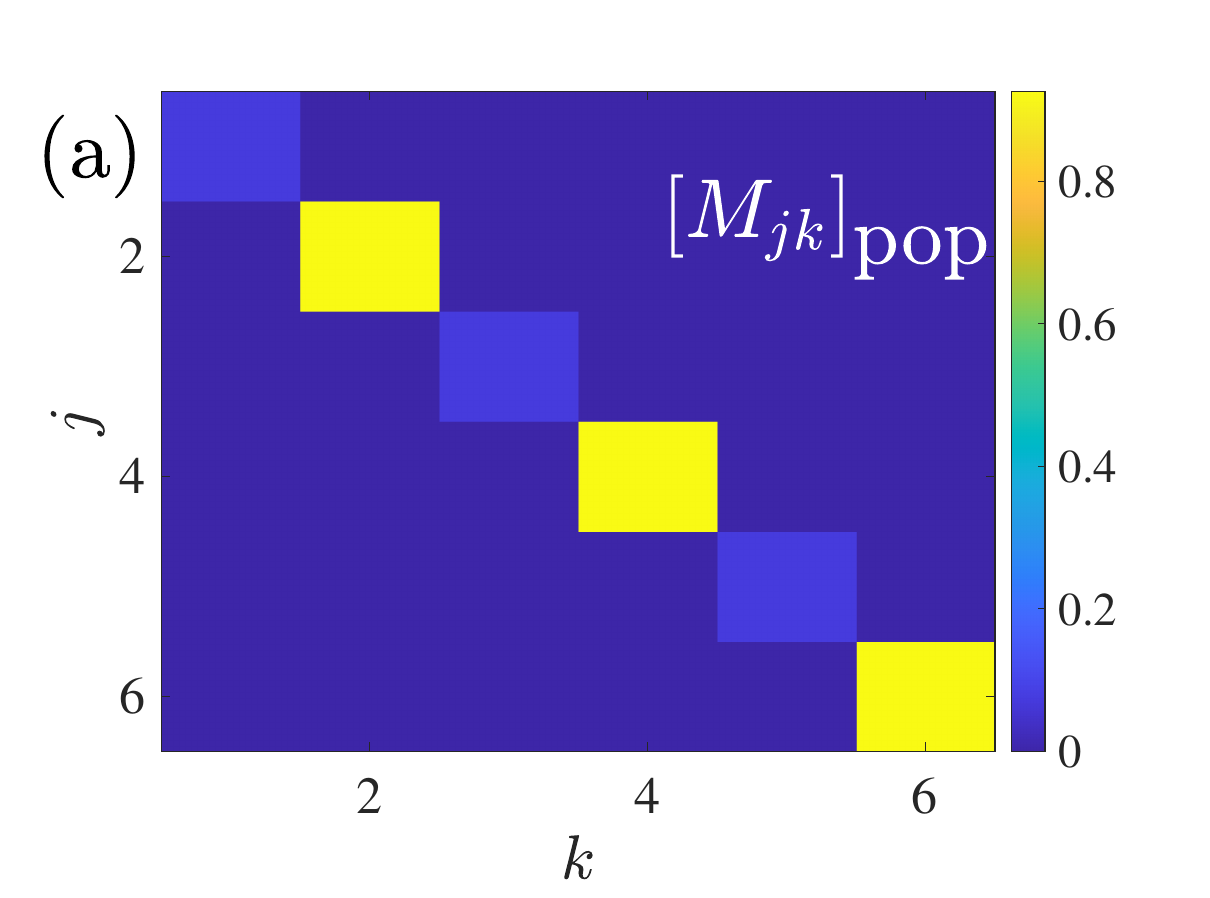}\hfill
\includegraphics[trim={0.25cm 7.5cm 0.25cm 7cm},clip, width=0.5\textwidth]{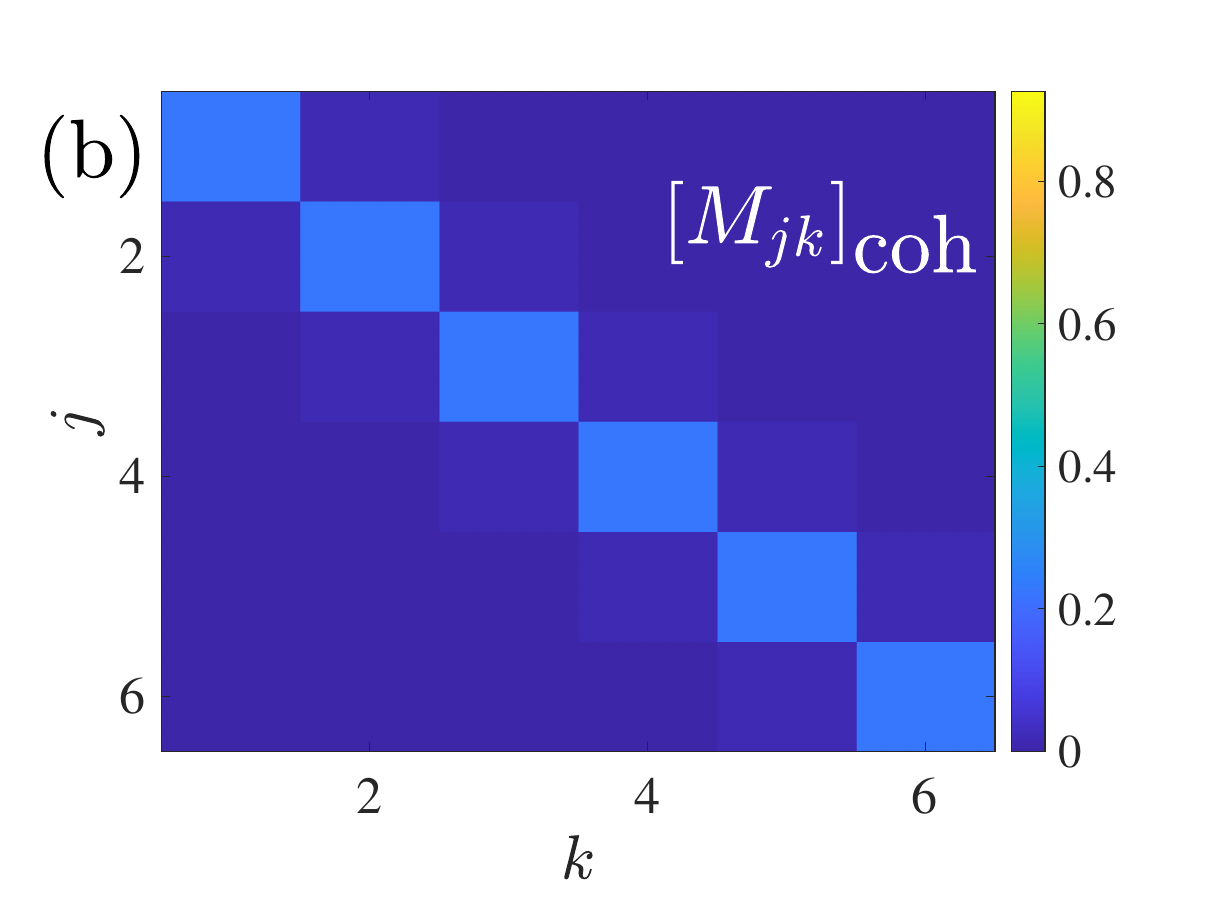}
\caption{Entries of $M_{jk}$. On the left, we have $\hat{\mathcal{M}}_\textrm{pop}$ when the probe wavelength is twice the lattice one and there is no phase difference between the probe and the optical lattice.  A $\pi/2$ phase shift between the probe and the lattice, while keeping the periodicity the same, leads to $\hat{\mathcal{M}}_\mathrm{coh}$ on the right.  The lattice depth equals five recoil energies.}\label{measuremat}
\end{figure*}

We write a dispersive measurement operator as
\begin{equation}
\hat{\mathcal{M}}_0 = \sum_{j,k} M_{jk} \hat{b}_{j}^\dagger \hat{b}_{k},
\label{M_Op_Expan}
\end{equation}
where $\hat{b}_{i}^\dagger$ creates a boson at the $i$th optical lattice site. One obtains the matrix elements in terms of the Wannier functions as
\begin{equation}
\label{MInt}
M_{jk}=\frac{g^2}{\Delta}\int|f_a(x,\omega_L)|^2w_j^*(x)w_k(x)\mathrm{d}x,
\end{equation}
where $g$ denotes the coupling strength between the probe laser and the ultracold atomic system, $f_a(x,\omega_L)$ is the spatial mode function, and $\Delta$ is the detuning of the probe from the atomic transition \cite{jaksch98,maschler,mekhov12}. For the two measurement operators considered in the main text,  the entries of the matrices are displayed as images in Fig.\ \ref{measuremat}.  From there we have
\begin{subequations}
\begin{align}
\label{MInt_Ex}
\hat{\mathcal{M}}_\mathrm{pop}:M_{jk} &\approx m_\mathrm{pop}\delta_{j,k}\delta_{\textrm{mod}(j,2),0}, \\
\hat{\mathcal{M}}_\mathrm{coh}:M_{jk} &\approx m_\mathrm{coh}\left(\delta_{j,k-1} + \delta_{j,k+1} \right) + d_\mathrm{coh}\delta_{j,k},
\end{align}
\end{subequations}
where $\textrm{mod}()$ denotes the modulo operation.  We ignore the term $d_\mathrm{coh}\delta_{j,k}$ in our numerical integration.  This is because this term leads to a constant shift $\hat{\mathcal{C}} = d_\mathrm{coh} N_{b}$ in $\hat{\mathcal{M}}_\mathrm{coh}$ with $N_{b} = 6$ being the total number of bosons in the system, and the normalized It\^{o} SSE
\begin{multline}
\text{d}|\psi(t)\rangle =\Bigl[-i\hat{\mathcal{H}} - \frac{\gamma}{2} \left(\hat{\mathcal{M}}_0 - \left\langle \hat{\mathcal{M}}_0 \right\rangle\right)^2 \text{d} t \\
+ \sqrt{\gamma}\left(\hat{\mathcal{M}}_0 - \left\langle \hat{\mathcal{M}}_0 \right\rangle \right)\text{d} W  \Bigr] |\psi(t)\rangle
\label{Supp_Eq:Homodyne_Ito_Normalized}
\end{multline}
remains unchanged under the transformation $\hat{\mathcal{M}}_0 \rightarrow \hat{\mathcal{M}}_0 + \hat{\mathcal{C}}$, where $\hat{\mathcal{C}}$ is a constant operator.

\section{Numerical Integration of the Stochastic Schr\"{o}dinger Equation}
\label{SecNumIntSSE}

We start with the It\^{o} SSE
\begin{equation}
\text{d}|\bar{\psi}(t)\rangle =\Bigl[-i\hat{\mathcal{H}} - \frac{\gamma}{2} \hat{\mathcal{M}}_0^2 + I(t) \hat{\mathcal{M}}_0 \Bigr]\text{d}t|\bar{\psi}(t)\rangle \label{Supp_Eq:Homodyne_Ito}
\end{equation}
that describes the time evolution of a non-normalized wavefunction  $|\bar{\psi}\rangle$.  We have written the homodyne measurement signal $I(t)$ as
\begin{equation}
I(t) = 2\gamma \langle \hat{\mathcal{M}_0} \rangle + \sqrt{\gamma} \: \text{d}W/\text{d}t.
\label{Supp_Eq:Homomeasure}
\end{equation}
To obtain the PSDs in Figs.\ 1(b,d,f,h) of the main text, we use the Stratonovich form of the SSE \cite{wiseman2009quantum, suppSDE}
\begin{equation}
\text{d}|\bar{\psi}(t)\rangle =\Bigl[-i\hat{\mathcal{H}} - \gamma \hat{\mathcal{M}}_0^2 + I(t) \hat{\mathcal{M}}_0 \Bigr]\text{d}t|\bar{\psi}(t)\rangle.\label{Supp_Eq:Homodyne_Strato}
\end{equation}
We need this form because the chain rule for Stratonovich equations is equivalent to the chain rule of conventional calculus.

Discretizing the full time interval $(0, t_\text{fin}]$, we write the wavefunction at the $(j+1)$th step as
\begin{equation}
|\bar{\psi}(t_{j+1})\rangle \approx  \left | \psi(t_{j}) \right\rangle + e^{\hat{\mathbb{G}} (t_j)} \left | \psi(t_{j}) \right\rangle,
\label{Supp_Eq:Krilov1}
\end{equation}
where
\begin{multline}
\hat{\mathbb{G}} (t_j) = \Bigl[ 1 -i\hat{\mathcal{H}} \delta t \\
+ \gamma\left(2\hat{\mathcal{M}_0} \left\langle \psi(t_{j}) \right | \hat{\mathcal{M}_0} \left | \psi(t_{j}) \right\rangle  - \hat{\mathcal{M}}_0^2  \right)\delta t \\
+ \sqrt{\gamma}\hat{\mathcal{M}}_0 \sqrt{\delta t}S_{j} \Bigr],
\label{Supp_Eq:Krilov2}
\end{multline}
$\delta t = t_{j+1} - t_{j}$ is the infinitesimal time increment, and $S_j$ is a random number drawn from a standard normal distribution.  To compute $e^{\hat{\mathbb{G}} (t_j)} \left | \psi(t_{j}) \right\rangle$, we use a Krylov subspace projection technique. Instead of computing the matrix exponential in isolation, this technique directly computes the action of the exponential operator on the wavefunction.  Although we used the normalized wavefunction $\left | \psi(t_{j}) \right\rangle$ on the right-hand side of Eq.\ \eqref{Supp_Eq:Krilov1}, we need to normalize the wavefunction again at the $(j+1)$th step using $\left | \psi(t_{j+1}) \right\rangle = |\bar{\psi}(t_{j+1})\rangle/ \sqrt{\langle \bar{\psi}(t_{j+1}) | \bar{\psi}(t_{j+1}) \rangle}$.

\begin{table}[ht]
\begin{center}
\begin{tabular}{ c  c  c  c }
\hline
\hline
Model & $t_\text{in}$ & $t_\text{fin}$ & $l$ \\
\hline

Bose-Hubbard & $1.9 \times 10^{5}$ & $2.0 \times 10^{5}$ & $50$ \\

Transverse-field Ising & $4.0 \times 10^{4}$ & $5.0 \times 10^{4}$ & $50$ \\
\hline

\end{tabular}
\end{center}
\caption{Numerical values of $t_\text{in}, t_\text{fin}$ and $l$.}
\label{Tab_Num_Process}
\end{table}

\begin{figure*}
\includegraphics[trim={0.25cm 3cm 0.25cm 9cm},clip, width=0.33\textwidth]{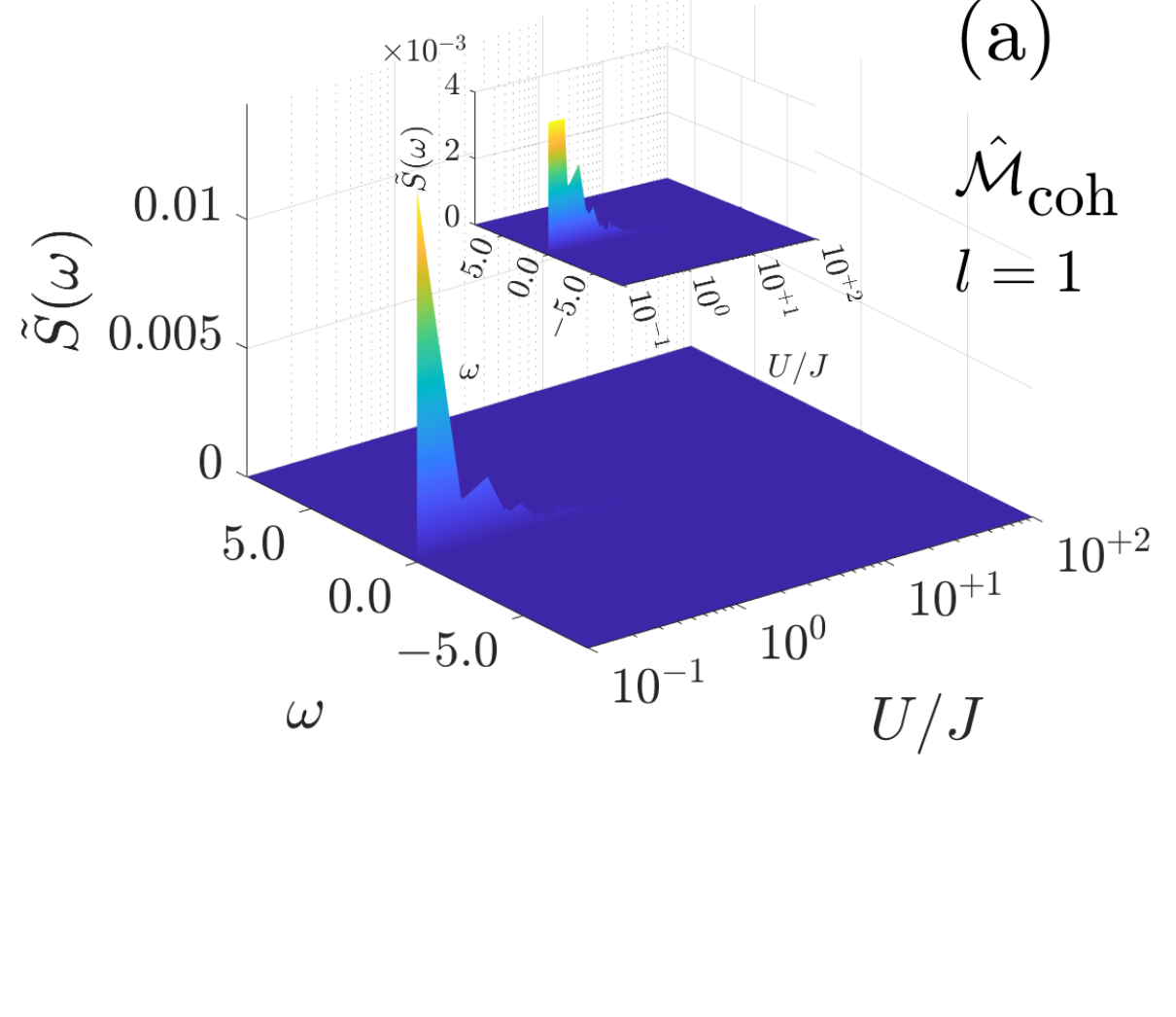}\hfill
\includegraphics[trim={0.25cm 3cm 0.25cm 9cm},clip, width=0.33\textwidth]{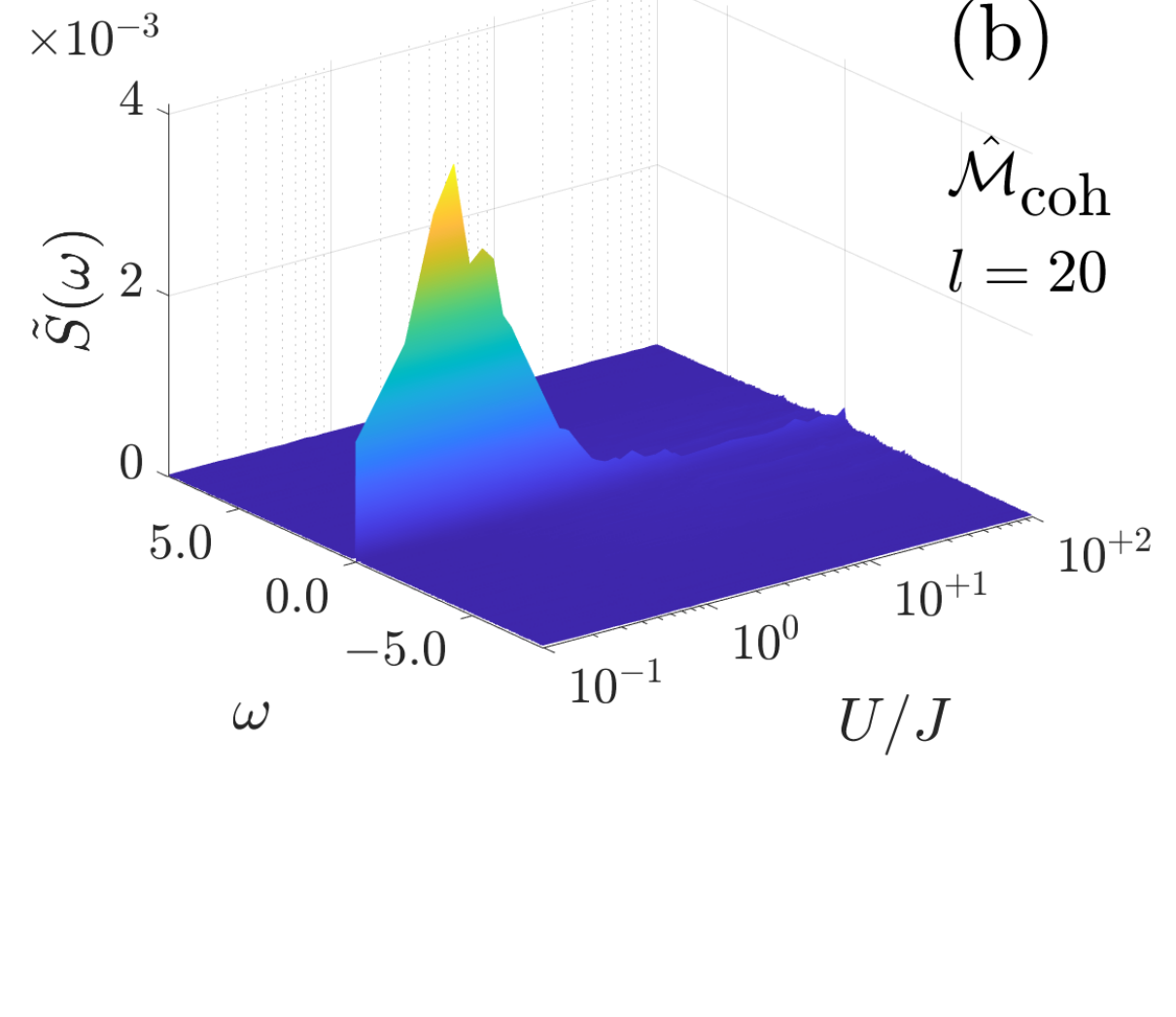}\hfill
\includegraphics[trim={0.25cm 3cm 0.25cm 9cm},clip, width=0.33\textwidth]{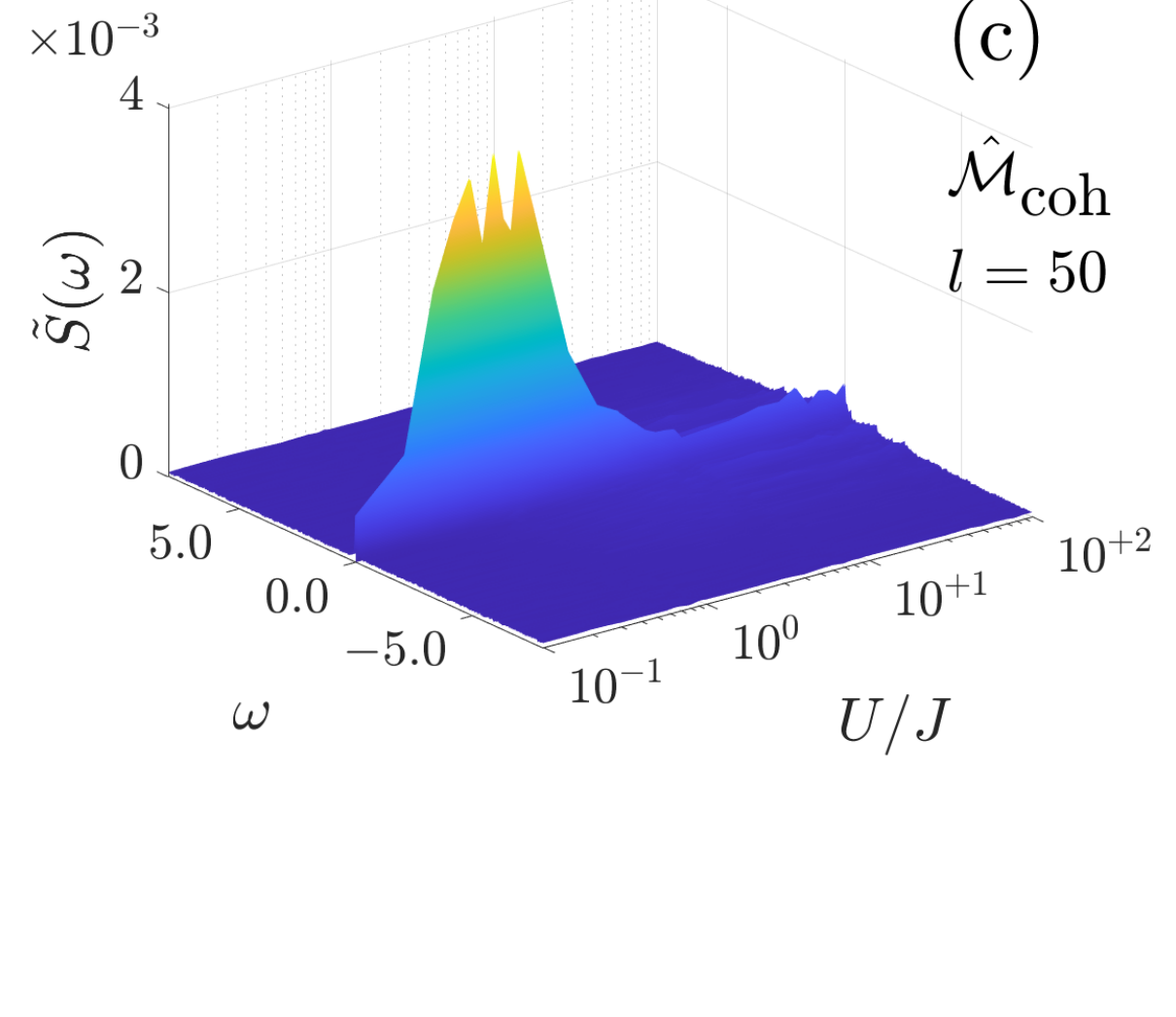}
\vspace{-1.0cm}
\caption{The dependence of a PSD on the noise averaging. We show the PSDs for the Bose-Hubbard model, where we measure $\hat{\mathcal{M}}_\mathrm{coh}$. The value of $l$ is $1, 20$ and $50$ in \textbf{(a), (b)} and \textbf{(c)}, respectively.  In the inset of \textbf{(a)}, we restrict the $\tilde{S}(\omega)$ range to $[0, 4\times 10^{-3}]$ for the $l = 1$ PSD, to have a better comparison with the PSDs in panels \textbf{(b)} and \textbf{(c)}. All the PSDs predict similar values for the transition point.}
\label{Supp_Fig:Diff_l}
\end{figure*}

After numerically obtaining the trajectories $\lbrace \left | \psi(t_{j}) \right\rangle, I(t_j) \rbrace$ for all the time steps in the interval $(0, t_\text{fin}]$, we discard the initial transients corresponding to the part $(0, t_\text{in}]$.  To obtain a noise averaged smoother PSD, we divide the considered quantum trajectory into $l$ parts and calculate the average PSD. The values of $t_\text{in}, \;t_\text{fin}$ and $l$ for Bose-Hubbard and the transverse-field Ising model are given in Table  \ref{Tab_Num_Process}.  We have considered the infinitesimal time increment $\delta t$ to be $0.01$ for all the trajectories.  We also show how the PSD depend on $l$ in Fig.\ \ref{Supp_Fig:Diff_l}.

\end{document}